\documentclass[preprint,tightenlines,nofootinbib]{revtex4-1}
\usepackage{amsmath,amssymb}
\usepackage{color}
\usepackage[colorlinks,citecolor=blue,urlcolor=blue]{hyperref}
\usepackage[sort&compress]{natbib}
\providecommand{\email}[1]{\href{mailto:#1}{#1}}
\usepackage{bm} 
\IfFileExists{hypernat.sty}{\usepackage{hypernat}}

\newcommand{\qt}{\tilde{q}}
\newcommand{\rvec}{\bm{r}}
\newcommand{\ovec}{\bm{\omega}}
\newcommand{\eab}{\epsilon_{ab}}
\newcommand{\euab}{\epsilon^{ab}}

\newcommand{\Reals}{\mathbb{R}}
\newcommand{\calN}{\mathcal{N}}
\newcommand{\tp}{\theta^{+}}
\newcommand{\tm}{\theta^{-}}
\newcommand{\tpb}{\bar{\theta}^{+}}
\newcommand{\tmb}{\bar{\theta}^{-}}
\newcommand{\sqt}{\sqrt{2}\,} 
\newcommand{\Qt}{\widetilde{Q}}

\newcommand{\calH}{\mathcal{H}}
\newcommand{\calL}{\mathcal{L}}
\newcommand{\calA}{\mathcal{A}}
\newcommand{\calD}{\mathcal{D}}
\newcommand{\tC}{\widetilde{C}}
\newcommand{\tX}{\widetilde{X}}
\newcommand{\tJ}{\widetilde{J}}
\newcommand{\bbX}{\mathbb{X}}

\newcommand{\tht}{\tilde{\theta}}
\newcommand{\tl}{\widetilde{\lambda}}
\newcommand{\tG}{\widetilde{G}}

\newcommand{\ra}{r_{\text{avg}}}
\newcommand{\rrel}{r_{\text{rel}}}

\newcommand{\form}[1]{\mathsf{#1}}

\begin{document}

\title{The {KK}-Monopole/{NS}5-Brane in Doubled Geometry}

\author{Steuard Jensen}
\email{jensens@alma.edu}
\affiliation{Department of Physics, 
Alma College\\614 W.\ Superior St., Alma, MI 48801}

\date{June 6, 2011}

\begin{abstract}
The Kaluza-Klein monopole has been recognized as a string background with significant non-geometric features: it must appear ``localized'' to winding strings to match the NS5-brane's localization on the T-dual circle. In this work, we explicitly construct this T-dual system in the doubled geometry formalism, which proves to successfully describe the duality despite a broken isometry on one side of the duality pair. We further suggest an extension of the doubled formalism to the gauged linear sigma models describing this system (both bosonic and supersymmetric) and show that previous calculations of worldsheet instanton effects are best understood in this doubled form.
\end{abstract}

\maketitle

\section{Introduction}

Most formulations of T-duality in string theory are based on the assumption that the full background must have an isometry: the metric, torsion, and dilaton must all be independent of the duality direction (e.g.~\cite{Buscher:1987qj,Rocek:1991ps}); the dual background necessarily has an isometry as well. This is a broad class of backgrounds, and the resulting duality web has taught us a great deal about the structure of string theory. However, there is considerable evidence that this story must be incomplete: requiring an isometry leaves unsatisfying gaps in the duality web, and even small perturbations seemingly render dualities invalid. But the tools available to study these cases are limited: among other issues, it is clear that the dual backgrounds cannot be interpreted geometrically in terms of a supergravity approximation.

The present discussion will focus on a specific example of such a relationship: the T-duality between the Kaluza-Klein monopole and the NS5-brane. It has long been known that the KK-monopole described by Gross and Perry~\cite{Gross:1983hb} and by Sorkin~\cite{Sorkin:1983ns} is a natural object in string theory. T-duality on its nontrivially fibered circle maps it to an $H$-monopole~\cite{Banks:1988rj,Ooguri:1995wj}, understood as an NS5-brane smeared around the circle~\cite{Gauntlett:1992nn,Khuri:1992ww}. (The T-duality of their world-volume actions was considered in~\cite{Eyras:1998hn}.)

However, the proper NS5-brane solution on a circle is localized at a point~\cite{Gauntlett:1992nn} that plays a significant role in the background geometry: strings with momentum around the circle would see a throat as they approached that position (at least when multiple NS5-branes are present). Based on the equivalence of physics across T-duality, Gregory, Harvey, and Moore argued that in string theory the Kaluza-Klein monopole solution should be modified as well~\cite{Gregory:1997te}: in a proper KK-monopole, winding strings should be able to see a corresponding throat behavior.

This conjecture was verified by Harvey and the present author by studying worldsheet instanton corrections to a gauged linear sigma model for the usual Kaluza-Klein monopole~\cite{Harvey:2005ab}. The argument used closely paralleled Tong's work on localization of the NS5-brane~\cite{Tong:2002rq}: instanton terms give corrections to the low energy effective action that can be interpreted as changes to the effective geometry of the solution. For the NS5-brane, this turns the smeared geometry into the known localized geometry. For the KK-monopole, the resulting solution is new and harder to interpret: it depends explicitly on the dual coordinate, and corresponds to localization in ``winding space.'' However, the authors of~\cite{Harvey:2005ab} were unable to provide a mathematical formalism that could simultaneously describe a geometric coordinate and its (non-geometric) T-dual.

``Doubled geometry,'' as formulated by Hull~\cite{Hull:2004in,Hull:2006va}, is precisely such a formalism. In it, the target space is enlarged: for each dimension where T-duality will act the dual dimension is added to the manifold, leading to an action in which T-duality is manifest. Constraints are then imposed to remove the extra degrees of freedom. Although most discussions of this formalism assume that all fields are independent of the duality coordinates, this does not appear to be a fundamental requirement. Dabholkar and Hull have shown how doubled geometry can describe the dependence on such coordinates that arises naturally when T-duality acts on backgrounds with fluxes~\cite{Dabholkar:2005ve}. In discussing their results, they comment that the ``winding localized'' KK-monopole is another such background.

The present work makes this realization of the Kaluza-Klein monopole in doubled geometry explicit. This provides the first full description of this object in string theory that includes the details of its localized behavior as seen by winding strings. It is surprising to find that such a familiar object requires a new formalism such as doubled geometry for its complete description, but the only alternative used previously was to attribute these features to a loosely characterized coherent state of winding modes in string field theory (as in~\cite{Gregory:1997te}). The success of the doubled formalism in reproducing previous results in this specific case provides further evidence that the formalism can give valuable insight in general backgrounds without isometries.

The plan of this paper is as follows. In section~\ref{sec:Conventional} we review the geometries of the NS5-brane and the Kaluza-Klein monopole in both their smeared and localized forms, including a summary of previous results for the ``winding localized'' KK-monopole. In sections~\ref{sec:DoubledSmeared} and~\ref{sec:DoubledLocalized} we review the doubled geometry formalism and apply it to the KK-monopole/NS5-brane system, arguing that it remains consistent despite the broken isometry on one side of the duality. Section~\ref{sec:GLSM} extends the doubled formalism to both bosonic and supersymmetric gauged linear sigma models for these systems, leading in section~\ref{sec:localization} to a brief review of the instanton calculation of~\cite{Harvey:2005ab}, which is clarified in this improved context. Finally, in section~\ref{sec:Conclusions} we discuss the implications of these results and directions for further study.

\section{Conventional and doubled monopole backgrounds}
\label{sec:MonopoleBkgds}
\subsection{Conventional formulation}
\label{sec:Conventional}

\subsubsection{The NS5-brane background}
\label{sec:ConventionalNS5}

The supergravity background associated with the NS5-brane has the same basic structure in both its smeared and localized forms. In the four dimensions where the fields are nontrivial the metric and torsion are given in the conventions of~\cite{Harvey:2005ab} by
\begin{equation}
  \label{eq:NS5Geometry}
  ds^2 = H(r,\theta) \left( d\rvec \cdot d\rvec + d\theta^2 \right)
  \qquad \text{and} \qquad
  H_{\mu \nu \rho} = \epsilon_{\mu \nu \rho}{}^\lambda
    \partial_\lambda \ln H(r,\theta) \quad.
\end{equation}
Here, $\rvec = r^m$ is a position in $\Reals^3$ ($m,n,\dotsc=1,2,3$), $r^\mu$ also includes $\theta=r^4$ with period $2 \pi$, and $H(r,\theta)$ is a harmonic function equal in the smeared and localized cases to
\begin{equation}
  \label{eq:HarmonicFns}
  H(r,\theta) = H(r) =
  \frac{1}{g^2} + \frac{1}{2 r}
  \qquad \text{and} \qquad
  H(r,\theta) = \frac{1}{g^2} + \frac{1}{2r}
 \frac{\sinh r}{\cosh r - \cos \theta}
 \quad,
\end{equation}
respectively. The dilaton is given by $e^\Phi = H(r,\theta)$ in either case. The localized harmonic function becomes smeared when averaged over $\theta$ or in the large $r$ limit.

The torsion can be expressed in terms of a $B$-field as $\form{H}=\form{dB}$. In terms of spherical coordinates $\{r, \vartheta, \varphi\}$ for $\Reals^3$ (note that $\vartheta \ne \theta$), one common gauge choice yields just two independent non-vanishing components of $\form{B}$:\footnote{Here and throughout, we correct a minor error of~\cite{Harvey:2005ab} by reversing the signs of $\form{B}$ and $\omega_m$.}
\begin{equation}
\label{eq:BfieldNS5}
  B_{\varphi 4} = -r^2 (1 - \cos \vartheta) \partial_r H(r,\theta)
  \qquad \text{and} \qquad
  B_{r \varphi} = -r^2 (1 - \cos \vartheta) \partial_\theta
  H(r,\theta)
  \quad.
\end{equation}
In any gauge, after dimensional reduction the components $B_{m4}$ form  a vector potential $\ovec$ in $\Reals^3$: $\omega_m \equiv -2 B_{m4}$.

In the smeared case (or at large $r$), the $B$-field components given above become simply $B_{\varphi 4} = \frac{1}{2} (1 - \cos \vartheta)$ and $B_{r \varphi} = 0$. The corresponding $\ovec$ can be recognized as a magnetic  monopole potential for $H_{mnp}$ (an $H$-monopole): the only non-zero  component of $\ovec$ is $\omega_\varphi = -(1 - \cos \vartheta)$, so $\nabla \times \ovec = \nabla (1/r)$.

\subsubsection{The KK-monopole background}
\label{sec:ConventionalKK}

The Kaluza-Klein monopole was first derived as a solution of pure general relativity in five dimensions, described completely by the metric of Taub-NUT space (plus time)~\cite{Gross:1983hb,Sorkin:1983ns}. In the same conventions used above, this can be written in terms of the (smeared) harmonic function $H(r)$ as
\begin{equation}
  \label{eq:KKmonmetric}
  ds^2 = H(r)\, d \rvec \cdot d \rvec + H(r)^{-1}
  \bigl( d \tht + \tfrac{1}{2} \ovec \cdot d \rvec \bigr)^2\quad;
\end{equation}
the fourth coordinate $\tht=r^4$ has periodicity $2 \pi$. (The torsion and dilaton vanish, as this is a pure gravity solution.)  This solution is flat at the origin and has global topology $\Reals^4$: its local $\Reals^3 \times S^1$ structure is the Hopf fibration of $S^1$ over $S^2$ (times the radial coordinate $\Reals_+$). After dimensional reduction along $\tht$, $\ovec$ gives the vector potential of a magnetic monopole for the Kaluza-Klein gauge field: $\nabla \times \ovec = \nabla (1/r)$, and in an appropriate gauge the only non-vanishing component is again $\omega_\varphi = -(1-\cos \vartheta)$.

In pure gravity, this is the full solution; there is no ``localized'' case analogous to the NS5-brane. When embedded in string theory, Sen showed that this solution gains a dyonic degree of freedom from a large gauge transformation~\cite{Sen:1997zb}: $\form{B} = -\beta\, \form{d} [ \frac{1}{g^2 H(r)} (\form{d\tht} + \tfrac{1}{2} \omega_m \form{dr}^m) ]$ is the self-dual harmonic 2-form of Taub-NUT space multiplied by a dyonic parameter $\beta$. This parameter provides the KK-monopole with the correct number of moduli, and when time dependent it can carry string winding charge. However, its only physical contribution is a phase: the solution still does not show any features corresponding to the NS5-brane throat.

Thus, the true T-dual to the (localized) NS5-brane must involve fields beyond the usual geometric content of supergravity. Gregory, Harvey, and Moore argued in~\cite{Gregory:1997te} that from the perspective of string field theory this should involve a nontrivial background of winding modes. Some of these effects were computed explicitly by Harvey and the present author in~\cite{Harvey:2005ab}, where the leading corrections to the curvature and torsion for the localized KK-monopole solution were found in a worldsheet instanton calculation.

That calculation is reviewed in section~\ref{sec:localization}, but it is useful to quote the results here. The instanton effects were interpreted as corrections to the target space curvature and used to deduce corrections to the metric and torsion. From that perspective, in the localized KK-monopole the first term in Eq.~\eqref{eq:KKmonmetric} becomes simply $H(r,\theta)\, d\rvec \cdot d\rvec$: the smeared harmonic function is replaced with the localized version. This is a highly unusual result, as the metric depends on the non-geometric coordinate $\theta$ from the T-dual space. Such a result has no meaning in conventional geometry, but as discussed in section~\ref{sec:localization} it proves to be entirely appropriate in the context of doubled geometry.

The limits used in~\cite{Harvey:2005ab} precluded direct calculation of the other metric components, but it was conjectured there that the harmonic function (and $\ovec$) would experience corresponding changes throughout Eq.~\eqref{eq:KKmonmetric}:
\begin{equation}
  \label{eq:locKKmetric}
  ds^2 = H(r, \theta)\, d \rvec \cdot d \rvec +
  H(r, \theta)^{-1} \bigl( d \tht + \tfrac{1}{2} \ovec \cdot
    d \rvec \bigr)^2\quad,
\end{equation}
where $\ovec$ here depends on $\theta$ and matches the result from the localized NS5-brane (that is, $\omega_\varphi = 2 r^2 (1-\cos \vartheta) \partial_r H(r, \theta)$ in the usual gauge). The other result of~\cite{Harvey:2005ab} was a non-vanishing torsion: in spherical coordinates, $H_{r\vartheta\varphi} = r^2 \sin \vartheta\, \partial_\theta H(r, \theta)$ with all other components zero in the relevant limits.

It is interesting to note that these results are precisely what one would obtain by naively applying the Buscher rules for T-duality~\cite{Buscher:1987qj} to the localized NS5-brane background despite the broken isometry around the $\theta$-circle. This connection was not recognized in~\cite{Harvey:2005ab}, but as discussed in section~\ref{sec:DoubledLocalized} the result arises naturally in the doubled geometry formalism.

\subsubsection{Nonlinear sigma models}
\label{sec:NLSMconventions}

With the exception of the localized KK-monopole, these backgrounds can be expressed in the usual manner as nonlinear sigma models on the string worldsheet. As usual, the action is $S = \frac{1}{2\pi} \int \form{L} = \frac{1}{2\pi} \int d^2\sigma \calL$ where the bosonic terms of the Lagrangian
take the form~\cite{Howe:1985pm}
\begin{equation}
\begin{aligned}
  \form{L} &=
   -\frac{1}{2} G_{\mu \nu} \form{d}r^\mu \wedge *\form{d}r^\nu +
    \frac{1}{2} B_{\mu \nu} \form{d}r^\mu \wedge \form{d}r^\nu \quad,
  \text{ or as a Lagrangian density,} \\
  \calL &= -\frac{1}{2} \eta^{ab} G_{\mu \nu} \partial_a
  r^\mu \partial_b r^\nu - \frac{1}{2} \euab B_{\mu
    \nu} \partial_a r^\mu \partial_b r^\nu \quad.
\end{aligned}
\end{equation}
The relationship between form and component language serves to establish our conventions: $\sigma^{a,b,\ldots}=\tau,\sigma$ are coordinates on the worldsheet, $\eta^{ab} = (\begin{smallmatrix} -1&0\\0&1 \end{smallmatrix})$ is the flat worldsheet metric, and $\epsilon_{01}=1$. The spacetime metric and $B$-field for each smeared solution indicate that these can each be viewed as an $S^1$ bundle over a base space with coordinates $r^m$:
\begin{alignat}{2}
  \label{eq:sNS5nlsm}
  \calL_\text{NS5 (smeared)} &=  - \frac{1}{2} H(r) (\partial_a \theta)^2 
  - \frac{1}{2} \euab \omega_m \partial_a \theta \partial_b r^m
  &&- \frac{1}{2} H(r) (\partial_a r^m)^2
  \quad \text{and}\\
  \label{eq:sKKnlsm}
  \calL_\text{KK (smeared)} &= \;\;\; - \frac{1}{2} H(r)^{-1} 
  (\partial_a \tht + \tfrac{1}{2} \omega_m \partial_a r^m)^2
  &&- \frac{1}{2} H(r) (\partial_a r^m)^2
  \quad.
\end{alignat}
Including the KK-dyon mode in Eq.~\eqref{eq:sKKnlsm} simply requires adding the term $\beta\, \epsilon^{ab} \partial_a [ \frac{1}{g^2 H(r)} (\partial_b \tht + \tfrac{1}{2} \omega_m \partial_b r^m) ]$. This term is a total derivative when $\beta$ is constant, but even then it may be topologically significant.

\subsection{The smeared KK-monopole/NS5-brane in doubled geometry}
\label{sec:DoubledSmeared}

\subsubsection{The doubled monopole action}
\label{sec:DoubledSmearedAction}

Doubled geometry in its current form was introduced by Hull in~\cite{Hull:2004in}: by making the T-duality of string theory manifest one is able to describe non-geometric string backgrounds not accessible in conventional formulations. The search for a duality symmetric formulation of string theory has a long pedigree, including fundamental work by Duff~\cite{Duff:1989tf}, Tseytlin~\cite{Tseytlin:1990nb,Tseytlin:1990va}, Maharana and Schwarz~\cite{Maharana:1992my}, Siegel~\cite{Siegel:1993th}, and others; Hull's approach is closely related to the formalism of Cremmer, Julia, Lu, and Pope~\cite{Cremmer:1997ct}. In this section we summarize the essential features of Hull's doubled geometry by relating the specific example of the (smeared) Kaluza-Klein monopole/NS5-brane system to the general formalism as expressed in~\cite{Hull:2006va}.

The central concept of doubled geometry is that the target space of the worldsheet action is enlarged to include not just the usual geometric coordinates but their duals as well. More precisely, if the conventional target space for some string background is a torus fibration $T^d$ (over some base) with $d$ torus coordinates $X^i$, the doubled space instead has fibers $T^{2d}$ with $2d$ local coordinates $X^i$ and $\tX_i$. To maintain the correct number of degrees of freedom, a constraint is imposed to essentially require half of the $2d$ torus coordinate fields to be left-moving and the other half to be right-moving. This doubled space can be related to a conventional description by choosing a ``polarization'' in each coordinate patch: a choice of which half of the $T^{2d}$ coordinates are geometric and which are not. Different choices of polarization yield T-dual backgrounds.

We consider only the smeared KK-monopole/NS5-brane system, where there is just one torus coordinate ($d=1$) over a base $\Reals^3$ (suppressing the trivial $\Reals^{5,1}$) and the doubled Lagrangian takes the form
\begin{equation}
\label{eq:LDoubled}
  \calL_\text{d} = -\frac{1}{4} H(r) (\partial_a \theta)^2
  - \frac{1}{4} H(r)^{-1} (\partial_a \tht + \tfrac{1}{2} \omega_m
  \partial_a r^m)^2 - \frac{1}{4} \euab \omega_m \partial_a
  \theta \partial_b r^m - \frac{1}{2} H(r) (\partial_a r^m)^2 \quad.
\end{equation}
(It is interesting to observe that here and throughout this work, the doubled Lagrangian is mathematically the mean of those for the two individual polarizations.)  The self-duality constraint that enforces the proper number of degrees of freedom can be expressed as the vanishing of a conserved 1-form current $J = H(r)^{-1} (d\tht + \tfrac{1}{2} \omega_m dr^m) - *d\theta$, or in components,
\begin{equation}
\label{eq:Jsmeared}
  J_a = H(r)^{-1} (\partial_a \tht + \tfrac{1}{2} \omega_m \partial_a
  r^m) + \eab \partial^b \theta \quad.
\end{equation}
We can see that $J_a$ is conserved using the $\tht$ field equation: $\frac{1}{2} \partial^a [H(r)^{-1} (\partial_a \tht + \tfrac{1}{2} \omega_m \partial_a r^m)] = 0$. Thus, $J_a$ is the Noether current for the transformation $\delta \tht = \tilde{\alpha}$, plus a trivially conserved term (to be justified shortly). The constraint is then $J_a = 0$.

The corresponding conserved current for the transformation $\delta \theta = \alpha$ is
\begin{equation}
\label{eq:tJsmeared}
  \tJ_a = H(r) \partial_a \theta + \eab
  (\partial^b \tht + \tfrac{1}{2} \omega_m \partial^b r^m) \quad.
\end{equation}
Conservation of $\tJ_a$ results from the $\theta$ field equation: $\frac{1}{2} \partial^a[H(r) \partial_a \theta + \eab \tfrac{1}{2} \omega_m \partial^b r^m]=0$. Again, $\tJ_a$ also includes a trivially conserved term, but this time the reason for including it is clear: it is part of the familiar combination $(d \tht + \tfrac{1}{2} \ovec \cdot d \rvec)$ and required for manifest KK~gauge invariance. Also, once these terms are included $\tJ=H(r)\ {*J}$, and thus $\tJ=0$ is equivalent to $J=0$: it is possible to impose the self-duality constraint for the doubled geometry by forcing either current to vanish.\footnote{%
This example can be expressed in the general formalism of doubled geometry with a single doubled coordinate. In the language of~\cite{Hull:2004in, Hull:2006va}, the doubled Lagrangian is
\begin{equation*}
\label{eq:GeneralLDoubled}
  \calL_\text{d} = -\frac{1}{4} \eta^{ab} \calH_{IJ} (\partial_a
  \bbX^I + \calA^I_m \partial_a r^m) (\partial_b \bbX^J +
  \calA^J_n \partial_b r^n) - \frac{1}{2} \euab
  L_{IJ} \partial_a \bbX^I \calA^J_m \partial_b r^m - \calL(r^m)
  \quad.
\end{equation*}
The two fiber coordinates are collectively labeled $\bbX^I = (\theta, \tht)$ with positive definite fiber metric $\calH_{IJ}$ and connection 1-form components $\calA^I_a = \calA^I_m \partial_a r^m$. $L_{IJ} = (\begin{smallmatrix} 0&1\\1&0 \end{smallmatrix})$ is an $O(1,1)$ invariant metric on the fibers; its inverse is $L^{IJ}$. $\calL(r^m)$ gives the sigma model on the base space. The general form of the vanishing current constraint is
  $J^I_a = L^{IJ}\calH_{JK} (\partial_a \bbX^K + \calA^K_a)
    + \eab (\partial^b \bbX^I + \calA^{Ib}) = 0$. 
For the case under consideration,
$\calH_{IJ} = \left(\begin{smallmatrix} H(r) & 0\\ 0 & H(r)^{-1} \end{smallmatrix}\right)$, $\calA^I_m = (0, \tfrac{1}{2} \omega_m)$, and $J^I_a = (J_a, \tJ_a)$.}

The connection between this doubled action and the conventional formulations of the KK-monopole and NS5-brane comes when one chooses a polarization of the doubled space, identifying one of the fiber coordinates $\theta$ or $\tht$ as geometric and the other as a non-geometric dual. The constraint can then be used to eliminate (derivatives of) the dual coordinate from the action, which will reduce Eq.~\eqref{eq:LDoubled} to Eq.~\eqref{eq:sNS5nlsm} or~\eqref{eq:sKKnlsm}, respectively. The result is equivalent to the standard Buscher rules for T-duality~\cite{Buscher:1987qj}.

\subsubsection{Imposing the constraint}
\label{sec:DoubledSmearedConstraint}

Imposing the constraint by hand as described above makes its role clear, but this obscures the quantum effects of the choice of polarization. Instead, we impose the constraint within the action by gauging either the symmetry generated by $J$ or that generated by $\tJ$~\cite{Hull:2006va}. (Attempting to gauge both at once is inconsistent, essentially because the two constraints are equivalent). The choice of which to gauge is equivalent to a choice of polarization: one coordinate's dynamics are completely absorbed into the gauge freedom. Gauging is accomplished through minimal coupling of the non-geometric coordinate to a gauge field $C$ or $\tC$: $d\tht \to d\tht + C$ or $d\theta \to d\theta + \tC$. The total Lagrangian will then be the sum of $\calL_\text{d}$ from Eq.~\eqref{eq:LDoubled}, a topological term required for gauge invariance:
\begin{equation}
\label{eq:Ltop}
\calL_\text{top} = \frac{1}{2} \euab \partial_a
\theta \partial_b \tht \quad,
\end{equation}
and one of the following gauge Lagrangian terms (with linear coupling
to the conserved currents):
\begin{align}
  \calL_g &= -\frac{1}{2} \eta^{ab} C_a J_b - \frac{1}{4} H(r)^{-1}
  (C_a)^2 \nonumber\\
  \label{eq:gaugedJcurrent}
  &= -\frac{1}{4} H(r)^{-1} \left( (C_a)^2 + 2 \eta^{ab} C_a
    (\partial_b \tht + \tfrac{1}{2} \omega_m \partial_b r^m) \right) -
  \frac{1}{2} \euab C_a \partial_b \theta \qquad \text{or} \\
  \calL_{\tilde{g}} &= -\frac{1}{2} \eta^{ab} \tC_a \tJ_b -
  \frac{1}{4} H(r) (\tC_a)^2 \nonumber \\
  \label{eq:gaugedtJcurrent}
  &= -\frac{1}{4} H(r) \left( (\tC_a)^2 + 2 \eta^{ab} \tC_a \partial_b
    \theta \right) - \frac{1}{2} \euab \tC_a (\partial_b \tht
  + \tfrac{1}{2} \omega_m \partial_b r^m) \quad.
\end{align}

To see that the first gauging above does eliminate the non-geometric coordinate and reproduce the familiar sigma model for the NS5-brane, we simply change variables in the total Lagrangian $\calL_{\text{NS5}'} \equiv \calL_\text{d} + \calL_\text{top} + \calL_g$. Defining $D_a = C_a + H(r) J_a$ we find that
\begin{equation}
  \calL_{\text{NS5}'} = -\frac{1}{4} H(r)^{-1} (D_a)^2 - \frac{1}{2}
  H(r) (\partial_a \theta)^2 - \frac{1}{2} H(r) (\partial_a r^m)^2 -
  \frac{1}{2} \euab \omega_m \partial_a \theta \partial_b r^m
  \quad.
\end{equation}
This shows that the gauge field $D_a$ is a non-dynamical auxiliary field which at the classical level has no effect on the system. The rest of the Lagrangian is precisely that of the smeared NS5-brane background as given in Eq.~\eqref{eq:sNS5nlsm}.

Similarly, defining $\widetilde{D}_a = \tC_a + H(r)^{-1} \tJ_a$ we find that the total Lagrangian $\calL_{\text{KK}'} \equiv \calL_\text{d} + \calL_\text{top} + \calL_{\tilde{g}}$ can be written as
\begin{align}
\label{eq:LKKDsmeared}
  \calL_{\text{KK}'} &= -\frac{1}{4} H(r) (\widetilde{D}_a)^2 -
  \frac{1}{2} H(r)^{-1} (\partial_a \tht + \tfrac{1}{2}
  \omega_m \partial_a r^m)^2 - \frac{1}{2} H(r) (\partial_a r^m)^2 +
  \euab \partial_a \theta \partial_b \tht \quad.
\end{align}
This includes an auxiliary field term for $\tilde{D}_a$, the KK-monopole Lagrangian of Eq.~\eqref{eq:sKKnlsm}, and a purely topological term that depends on both the geometric coordinate and its dual. Although it is surprising to see the non-geometric coordinate $\theta$ still present in the action for a KK-monopole, the term where it appears is a total derivative and thus does not contribute to the classical equations of motion. We will see in section~\ref{sec:GLSMbos} that an analogous term serves to preserve invariance of the KK-monopole action under large gauge transformations in the gauged linear sigma model formulation of this system.

\subsubsection{Quantization and the dilaton}
\label{sec:DoubledSmearedQuantization}

As discussed in~\cite{Hull:2006va}, the doubled theory can be quantized after gauging. Different choices of polarization lead to the same quantum theory, so the end result is independent of which gauge Lagrangian is used. Gauge fixing then completely eliminates the non-geometric coordinate (including winding modes). The required ghost integral is trivial, and the determinant from integrating out the auxiliary field changes the functional measure for the base-space coordinates due to its dependence on the harmonic function $H(r)$ from the spacetime metric.

When we allow a curved worldsheet metric $h_{ab}$ to replace $\eta_{ab}$, we must also include the dilaton coupling
\begin{equation}
  \label{eq:DilCoupling}
  \calL_\text{dil} = \sqrt{h} \phi R \quad.
\end{equation}
Here, $R$ is the Ricci scalar on the worldsheet and $\phi$ is a scalar field on the doubled space $\Reals^3 \times T^2$. The change in functional measure from integrating over the auxiliary field $D_a$ for the NS5-brane or $\widetilde{D}_a$ for the KK-monopole gives a contribution at one loop corresponding to a shift in $\phi$. In the coupling of Eq.~\eqref{eq:DilCoupling}, $\phi$ is replaced by either
\begin{align}
  \Phi_\text{NS5} &= \phi + \frac{1}{2} \log H(r) \quad
  \text{or}\\
  \Phi_\text{KK}  &= \phi - \frac{1}{2} \log H(r) \quad,
\end{align}
respectively. This matches the usual one loop T-duality transformation of the dilaton $\Phi$ and is consistent with the known dilaton background in each case as long as the doubled space dilaton is
\begin{equation}
  \phi = \frac{1}{2} \log H(r) \quad.
\end{equation}

\subsection{The localized KK-monopole/NS5-brane in doubled geometry}
\label{sec:DoubledLocalized}

Doubled geometry was initially formulated only for the case of torus bundles: in particular, all fields are assumed to be independent of the coordinates on $T^{2d}$. However, there has been some progress in relaxing that constraint (see for example the discussion of ``doubled everything'' in~\cite{Hull:2006va}, much of~\cite{Dabholkar:2005ve}, and ``compactifications with $R$-flux'' in~\cite{Hull:2009sg}), so it is interesting to ask how the localized NS5-brane and KK-monopole could be embedded in the doubled formalism.

Based on the known and conjectured background fields given in section~\ref{sec:Conventional} for the two localized cases, the natural conjecture for the doubled Lagrangian is:
\begin{equation}
\label{eq:LdLoc}
\begin{split}
  \calL_\text{d} + \calL_\text{top} = &\mbox{}-\frac{1}{4} H(r,\theta)
  (\partial_a \theta)^2 - \frac{1}{4} H(r,\theta)^{-1} (\partial_a
  \tht + \tfrac{1}{2} \omega_m \partial_a r^m)^2 - \frac{1}{4}
  \euab \omega_m \partial_a \theta \partial_b r^m\\
  &\mbox{} - \frac{1}{2} H(r,\theta) (\partial_a r^m)^2 - \frac{1}{2}
  \euab B_{mn}\, \partial_a r^m \partial_b r^n + \frac{1}{2}
  \euab \partial_a \theta \partial_b \tht \quad.
\end{split}
\end{equation}
The non-vanishing components of the antisymmetric terms can be found explicitly (for the usual gauge) in Eq.~\eqref{eq:BfieldNS5}: $\omega_\varphi = 2 r^2 (1-\cos \vartheta) \partial_r H(r, \theta)$ and $B_{r\varphi} = -r^2 (1 - \cos \vartheta) \partial_\theta H(r,\theta)$. The dilaton coupling may also be present as in Eq.~\eqref{eq:DilCoupling}, and the natural conjecture for the form of the scalar field is $\phi = \frac{1}{2} \log H(r,\theta)$.

The current corresponding to shifts $\delta \tht = \tilde{\alpha}$ is
\begin{equation}
  \label{eq:localJ}
  J_a = H(r,\theta)^{-1} (\partial_a \tht + \tfrac{1}{2}
  \omega_m \partial_a r^m) + \eab \partial^b \theta \quad.
\end{equation}
It is conserved just as in the smeared case, even though there is no longer a conserved current $\tJ$ corresponding to shifts in $\theta$. This implies that it is still consistent to impose the constraint $J_a=0$ by gauging the corresponding symmetry. The algebra involved is formally identical to that in the previous section, leading to
\begin{equation}
\begin{split}
  \calL_{\text{NS5}'} =& -\frac{1}{4} H(r,\theta)^{-1} (D_a)^2 -
  \frac{1}{2} H(r,\theta) (\partial_a \theta)^2 - \frac{1}{2}
  H(r,\theta) (\partial_a r^m)^2\\
  &- \frac{1}{2} \euab \omega_m \partial_a \theta \partial_b
  r^m - \frac{1}{2} \euab B_{mn} \partial_a r^m \partial_b r^n
  \quad.
\end{split}
\end{equation}
Apart from the auxiliary field term this is exactly the worldsheet Lagrangian for the localized NS5-brane. Given the doubled space dilaton background $\phi$ conjectured above, the same methods for quantization of the doubled theory with this choice of polarization reproduce the conventional treatment of the NS5-brane. This procedure suggests that the doubled theory remains well-defined at the quantum level. Although it was necessary to choose a specific polarization to obtain this result, it seems reasonable to interpret it as a correct quantum description of the full doubled theory as in other examples from doubled geometry.

Because we cannot gauge $\tJ$ in the localized case, these methods are not applicable to the opposite choice of polarization with $\tht$ as the physical coordinate. This is not a surprise: the ``localization'' of the Kaluza-Klein monopole solution is invisible from the standpoint of conventional geometry and thus we would not expect to find a consistent reduction of the complete physics in that language. (Other approaches to quantization of the doubled theory do not require a choice of polarization~\cite{HackettJones:2006bp,Berman:2007vi,Berman:2007xn,Berman:2007yf}, but it is not clear how to generalize them in the absence of an isometry.)

However, at the classical level it is possible to use the valid constraint $J_a=0$ to make contact with the results of~\cite{Harvey:2005ab} for the KK-monopole case. As suggested in~\cite{Hull:2009sg}, even if it is not possible to completely eliminate the non-geometric coordinate $\theta$ from the Lagrangian it is at least possible to hide the dynamics resulting from $\partial_a \theta$:
\begin{equation}
\begin{split}
  \calL_{\text{KK}'} = &-\frac{1}{2} H(r,\theta) (\partial_a r^m)^2 -
  \frac{1}{2} H(r,\theta)^{-1} (\partial_a \tht + \tfrac{1}{2}
  \omega_m \partial_a r^m)^2\\
  &- \frac{1}{2} \euab B_{mn}\, \partial_a r^m \partial_b r^n
  + \euab \partial_a \theta \partial_b \tht \quad. 
\end{split}
\end{equation}
The topological term depending on $\partial_a \theta$ is still present in this Lagrangian just as in the smeared case (Eq.~\eqref{eq:LKKDsmeared}), but as noted there it does not contribute to the equations of motion and thus has no impact on the classical system. This Lagrangian is entirely consistent with the worldsheet instanton results of~\cite{Harvey:2005ab}.

This result implies the following complete background for the ``winding localized'' Kaluza-Klein monopole in the usual gauge:
\begin{equation}
\label{eq:KKlocBackground}
\begin{gathered}
  ds^2 = H(r,\theta)\, d \rvec \cdot d \rvec + H(r,\theta)^{-1}
  \bigl( d \tht + r^2 (1 - \cos \vartheta) \partial_r H(r,\theta)
    \, d\varphi \bigr)^2 \quad,\\
  B_{r\varphi} = -r^2 (1 - \cos \vartheta) \partial_\theta H(r,\theta)
  \quad,
\end{gathered}
\end{equation}
with all other independent $B_{mn}$ zero. This is precisely the solution one would obtain by naive application of the Buscher rules for T-duality to the localized NS5-brane background given in Eqs.~\eqref{eq:NS5Geometry} and~\eqref{eq:BfieldNS5}.

The Buscher rules have not generally been considered applicable when the background fields depend on the duality coordinate: not only does this violate the assumptions under which they were derived, but the resulting dual fields retain their dependence on that coordinate even though it is no longer geometrically relevant. While $\partial_a \theta$ does not appear in the action above, it is still not appropriate to view $\theta$ as a constant. Indeed, the solution~\eqref{eq:KKlocBackground} is essentially meaningless in conventional geometry, and as discussed in section~\ref{sec:Conclusions} pointlike probes are not sensitive to the resulting changes. Describing the system in this form may give insight into the physics of the Kaluza-Klein monopole winding mode background, but it seems likely that these issues will make direct quantization impossible for this polarization. Instead, any discussion of the complete Kaluza-Klein monopole solution including its winding mode background should be formulated in the language of doubled geometry.

\section{Gauged linear sigma models and the doubled formalism}
\label{sec:GLSM}

One of the most surprising aspects of the results above is that the instanton calculation of~\cite{Harvey:2005ab}, which was based on a conventional formulation of the Kaluza-Klein monopole action, led to a result that could only be properly understood in the context of doubled geometry. Below, we summarize that calculation and the earlier NS5-brane calculation of~\cite{Tong:2002rq}.

We first briefly review the bosonic gauged linear sigma model (GLSM) formulations of the NS5-brane and KK-monopole backgrounds and show how that formalism can be naturally extended to a doubled form. We then do the same for the models' natural $\calN = (4,4)$ supersymmetric extensions. Finally, we discuss the impact of worldsheet instantons and argue that earlier calculations are unchanged in the doubled picture and indeed may be best understood in that context. Throughout this section, the conventions used are those of~\cite{Harvey:2005ab} (with minor exceptions\footnote{The fields $\theta$, $\gamma$, $\tht$, $\tG'$, and $\Gamma$ differ from~\cite{Harvey:2005ab} by a sign as defined here, and $e^{i \alpha_\text{here}} = -i e^{-i \alpha_\text{there}}$.}).

\subsection{Bosonic gauged linear sigma models for monopoles}
\label{sec:GLSMbos}

\subsubsection{Conventional gauged linear sigma models}
\label{sec:bosGLSMconv}

The (smeared) NS5-brane background can be viewed as the low energy limit of a gauged linear sigma model on the worldsheet with the Lagrangian
\begin{equation}
\label{eq:NS5GLSMbos}
\begin{split}
  \calL_\text{NS5,GLSM} = \mbox{}&\frac{1}{2e^2} F_{01}^2
  -\frac{1}{2g^2} (\partial_a r^m)^2 - \frac{1}{2g^2} (\partial_a
  \theta)^2 - |\calD_a q|^2 - |\calD_a \qt|^2 + \theta
  F_{01}\\ 
  &\mbox{}- \frac{e^2}{2} \left( |q|^2 - |\qt|^2 - r^3 \right)^2 -
  \frac{e^2}{2} \left| 2 \qt q - \left( r^1 + i r^2 \right) \right|^2
   \quad.
\end{split}
\end{equation}
Here, the worldsheet coordinates are $\sigma^a$ (not $x^\mu$ as in~\cite{Harvey:2005ab}), $A_a$ is an abelian gauge field with field strength $F_{ab}$, $\theta$ and the $r^m$ are uncharged scalar fields, and $q$ and $\qt$ are oppositely charged scalars with covariant derivatives $\calD_a q = \partial_a q + i A_a q$ and $\calD_a \qt = \partial_a \qt - i A_a \qt$.

The gauge coupling $e$ is the only dimensionful parameter in the Lagrangian, so the low energy limit corresponds to $e \to \infty$. In this limit the gauge kinetic term vanishes leaving $A_a$ as an auxiliary field, and both of the potential terms on the second line of Eq.~\eqref{eq:NS5GLSMbos} must be zero in any finite energy solution. This constraint allows us to express the four real degrees of freedom in $q$ and $\qt$ in terms of the $r^m$ and the phase $\alpha$ of $q$:
\begin{equation}
\label{eq:qsoln}
q = \frac{1}{\sqrt{2}} e^{i \alpha} \sqrt{r + r^3} \quad,
\qquad
\qt = \frac{1}{\sqrt{2}} e^{-i \alpha}
  \frac{r^1 + i r^2}{\sqrt{r + r^3}} \quad.
\end{equation}
As before, $r = \sqrt{(r^m)^2}$, so $|q|^2+|\qt|^2=r$. Under a gauge transformation $A_a \to A_a + \partial_a \lambda$ the field $q$ transforms as $q \to e^{-i \lambda} q$, so $\alpha \to \alpha - \lambda$. When it is necessary to make an explicit choice of gauge, we will require real $q >0$: this corresponds to $\alpha=0$.

In the context of these low energy constraints, the target space vector $\omega_m$ is defined implicitly by $-i \left( q^\dag \partial_a q - q \partial_a q^\dag - \qt^\dag \partial_a \qt + \qt \partial_a \qt^\dag \right) \equiv 2r (\partial_a \alpha + \tfrac{1}{2} \omega_m \partial_a r^m)$. Fixing a gauge and integrating out the auxiliary vector leads to the additional relation $A_a = -(\partial_a \alpha + \frac{1}{2} \omega_m \partial_a r^m + \frac{1}{2r} \eab \partial^b \theta)$. Substituting these into the Lagrangian yields precisely the nonlinear sigma model for the smeared NS5-brane in Eq.~\eqref{eq:sNS5nlsm} (up to unimportant total derivatives). In $\alpha=0$ gauge the explicit form of Eq.~\eqref{eq:qsoln} implies the usual gauge for $\omega_m$, and sigma model gauge transformations that would shift $\alpha$ can be absorbed instead into target space gauge transformations of $\omega_m$.

\bigskip

The gauged linear sigma model Lagrangian for the T-dual KK-monopole background is:\footnote{This GLSM can be found from Eq.~\eqref{eq:NS5GLSMbos} by writing its $\theta$-dependent terms in first order form. First write $\calL$ in terms of $\partial \theta$, keeping total derivatives: $\mathcal{L}_\theta = -\frac{1}{2g^2} (\partial_a \theta)^2 +\theta F_{01} = -\frac{1}{2g^2} (\partial_a \theta)^2 + \euab \partial_a \theta \, A_b - \euab \partial_a (\theta A_b)$. Then replace $\partial_a \theta$ with a vector field $C_a$ and add a Lagrange multiplier $\gamma$: $\mathcal{L}_C = -\frac{1}{2g^2} (C_a)^2 + \euab C_a A_b - \euab \partial_a (\theta A_b) - \euab C_a \partial_b \gamma + \euab \partial_a \theta \partial_b \gamma$. The final term (a total derivative) ensures that this action reduces precisely to $\mathcal{L}_\theta$ when $\gamma$ is integrated out: the $\gamma$ equation of motion forces $C_a$ to be pure gauge, $C_a = \partial_a \theta$, and the $\gamma$ terms cancel. Including this term also makes $\mathcal{L}_C$ manifestly gauge invariant when written in terms of $(\partial_a \gamma - A_a)$. Integrating out $C_a$ instead of $\gamma$ produces the $\gamma$-dependent terms in Eq.~\eqref{eq:KKGLSMbos}.}
\begin{equation}
\label{eq:KKGLSMbos}
\begin{split}
  \calL_\text{KK,GLSM} = \mbox{}&\frac{1}{2e^2} F_{01}^2
  -\frac{1}{2g^2} (\partial_a r^m)^2 - \frac{g^2}{2} (\partial_a
  \gamma - A_a)^2 - |\calD_a q|^2 - |\calD_a \qt|^2\\
  &\mbox{}- \frac{e^2}{2} \left( |q|^2 - |\qt|^2 - r^3 \right)^2 -
  \frac{e^2}{2} \left| 2 \qt q - \left( r^1 + i r^2 \right) \right|^2
  + \euab \partial_a [ \theta (\partial_b \gamma - A_b)]\quad.
\end{split}
\end{equation}
The final term is a total derivative, but boundary terms are topologically significant in the instanton calculation outlined in section~\ref{sec:localization}. As before, it is unusual to see the dual field $\theta$ appear here, especially in a term that can lead to physically significant results. This is an indication that this calculation is best analyzed in a doubled formalism.

For Eq.~\eqref{eq:KKGLSMbos} to be gauge invariant, the gauge transformation of $\gamma$ must be a simple shift: when $A_a \to A_a + \partial_a \lambda$ then $\gamma \to \gamma + \lambda$. This suggests a natural gauge invariant combination of real scalars
\begin{equation}
\label{eq:thtDef}
\tht \equiv \gamma + \alpha \quad,
\end{equation}
which acts as the fourth coordinate field. The term $\euab \partial_a \theta \partial_b \gamma$ in Eq.~\eqref{eq:KKGLSMbos} ensures that the action is invariant even under large gauge transformations of $A_a$ for which the total derivative is significant.

Taking the low energy limit of this Lagrangian leads to the same constraints as in the NS5-brane case above. This time, the $A_a$ equation of motion gives $A_a = \partial_a \gamma - \frac{1}{g^2 H(r)} (\partial_a \tht + \frac{1}{2} \omega_m \partial_a r^m)$. The end result is precisely the nonlinear sigma model for the KK-monopole, Eq.~\eqref{eq:sKKnlsm}, plus the total derivative term $\euab \partial_a [ \theta \, \frac{1}{g^2 H(r)} (\partial_b \tht + \frac{1}{2} \omega_m \partial_b r^m)]$. Distributing the derivative, this can be written as $\theta\, \euab \partial_a [ \frac{1}{g^2 H(r)} (\partial_b \tht + \frac{1}{2} \omega_m \partial_b r^m)] + \euab \frac{1}{g^2 H(r)} \partial_a \theta (\partial_b \tht + \frac{1}{2} \omega_m \partial_b r^m)$. The first term is precisely the KK-dyon $B$-field with $\theta$ as the dyonic parameter. The second mixes in dynamics of the non-geometric coordinate which makes its interpretation unclear; this could be an artifact of the procedure used in deriving this duality, or it may be that a non-constant dyonic coordinate can only be properly understood within the doubled formalism. (It is interesting that the KK-dyon term arises naturally in this GLSM description of the system: when deriving the KK-monopole action from duality of an NSLM as in section~\ref{sec:MonopoleBkgds}, it is necessary to add it by hand.)

\subsubsection{Doubled formulation of the gauged linear sigma model}
\label{sec:bosGLSMdoub}

Because these gauged linear sigma models can be written in forms that have no dependence on either $\theta$ or $\gamma$ undifferentiated, they are reminiscent of the smeared monopoles' nonlinear sigma models and it is natural to ask how this could be promoted to a doubled formalism. (It may be premature to refer to this as doubled \emph{geometry}, since the GLSM does not have an immediate geometric interpretation.)  Just as in the nonlinear sigma model, the terms that do not involve $\partial_a \theta$ or $\partial_a \gamma$ are merely spectators in the mathematics of the doubled fields. The natural conjecture for a doubled GLSM is:
\begin{equation}
\label{eq:doubledGLSMbos}
\begin{split}
  \calL_\text{d,GLSM} = \mbox{}&\frac{1}{2e^2} F_{01}^2
  -\frac{1}{2g^2} (\partial_a r^m)^2 - \frac{1}{4g^2} (\partial_a
  \theta)^2 - \frac{g^2}{4} (\partial_a \gamma - A_a)^2
  - |\calD_a q|^2 - |\calD_a \qt|^2  + \frac{1}{2}
  \theta F_{01}\\
  &\mbox{}-\frac{e^2}{2} \left( |q|^2 - |\qt|^2 - r^3 \right)^2 -
  \frac{e^2}{2} \left| 2 \qt q - \left( r^1 + i r^2 \right) \right|^2
  + \frac{1}{2} \euab \partial_a [ \theta (\partial_b \gamma - A_b)]
  \quad.
\end{split}
\end{equation}
To relate this to the previous GLSMs it is necessary to choose a polarization and eliminate the non-physical coordinate by gauging as in section~\ref{sec:DoubledSmeared}. The calculations involved for this GLSM are almost identical to the smeared monopole case treated there, with $-A_a$ substituted for $\frac{1}{2} \omega_m \partial_a r^m$, $\frac{1}{g^2}$ for $H(r)$, and $\gamma$ for $\tht$. Depending on our choice of polarization, we add the gauge Lagrangian corresponding to either Eq.~\eqref{eq:gaugedJcurrent} or Eq.~\eqref{eq:gaugedtJcurrent} to $\calL_\text{d,GLSM}$. The topological term needed for the doubled formalism (Eq.~\eqref{eq:Ltop}) is already present in the GLSM: as we have seen, it arose naturally as a total derivative term.

Depending on choice of polarization, the end result is exactly the GLSM~\eqref{eq:NS5GLSMbos} for the NS5-brane plus the decoupled auxiliary term $-\frac{g^2}{4} (D_a)^2$ or the GLSM~\eqref{eq:KKGLSMbos} for the KK-monopole plus $-\frac{1}{4g^2} (\widetilde{D}_a)^2$. Thus, at least at the classical level this doubled gauged linear sigma model agrees with the conventional results.\footnote{One might hope to relate the low energy limit of Eq.~\eqref{eq:doubledGLSMbos} to the original doubled (smeared) monopole Lagrangian of Eq.~\eqref{eq:LDoubled} before choosing a polarization. However, the low energy limit involves integrating out $A_a$, which also appears in the gauge Lagrangian terms $\mathcal{L}_{g/\tilde{g}}$ added when choosing a GLSM polarization. It would be interesting to understand a sensible low energy limit before gauging.}

\subsection{Supersymmetric gauged linear sigma models for monopoles}
\label{sec:SUSYGLSM}

\subsubsection{Review of conventional calculations}
\label{sec:convGLSMsusy}

Instanton calculations such as those of~\cite{Tong:2002rq,Harvey:2005ab} require the full $\calN = (4,4)$ supersymmetric forms of these gauged linear sigma models, so we briefly review them before conjecturing an appropriate doubled form. We will write the $\calN=(4,4)$ supermultiplets as pairs of $\calN=(2,2)$ superfields (see~\cite{Witten:1993yc} for a review of this formalism). The defining components of each supermultiplet are given below; more details on the notation can be found in the references.

The actions for both the NS5-brane and the KK-monopole include a $\calN=(4,4)$ vector multiplet, which decomposes into an $\calN=(2,2)$ chiral superfield $\Phi$ and vector superfield $V$:
\begin{equation}
\label{eq:vectormult}
\begin{split}
\Phi & = \phi + \sqt \tp \tl_+ 
        + \sqt \tm \tl_-
	+ \sqt \tp \tm ( D^1 - i D^2 )
	+ \dotsb \\
V & = \tp \tpb A_+ + \tm \tmb A_-
        - \sqt \tm \tpb \sigma 
	- \sqt \tp \tmb \sigma^\dag\\
    & \quad \mbox{}
        - 2 i \tp \tm \left(\tmb \bar{\lambda}_- 
	                            + \tpb \bar{\lambda}_+ \right)
        - 2 i \tmb \tpb \left(\tp \lambda_+ 
                                      + \tm \lambda_- \right)
	+ 2 \tp \tm \tmb \tpb D^3 \\
\Sigma & \equiv \frac{1}{\sqrt{2}} \bar{D}_+ D_- V
      = \sigma + i \sqt \tp \bar{\lambda}_+
        - i \sqt \tmb \lambda_-
	+ \sqt \tp \tmb ( D^3 - i F_{01} )
	+ \dotsb \quad.
\end{split}
\end{equation}
The vector superfield is written in Wess-Zumino gauge and we have defined $A_\pm \equiv A_0 \pm A_1$; its gauge transformations are given by $V \to V + i(\Lambda - \Lambda^\dag)$ for arbitrary chiral superfield $\Lambda$. $\Sigma$ is a gauge invariant twisted chiral superfield containing the field strength $F_{01}$. The components of fermion doublets are labeled as $\lambda_\alpha = (\lambda_-, \lambda_+)$.

Both actions also include a charged hypermultiplet that decomposes into oppositely charged chiral superfields $Q$ and $\Qt$:
\begin{equation}
\label{eq:hypermult}
\begin{split}
Q & = q + \sqt \tp \psi_+ + \sqt \tm \psi_- + 2 \tp \tm F
        + \dotsb\\
\Qt & = \qt + \sqt \tp \tilde{\psi}_+ + \sqt \tm \tilde{\psi}_- 
        + 2 \tp \tm \widetilde{F}
	+ \dotsb \quad.
\end{split}
\end{equation}
The NS5-brane action includes a twisted hypermultiplet that decomposes into a chiral superfield $\Psi$ and a twisted chiral superfield $\Theta$:
\begin{equation}
\label{eq:twistedmult}
\begin{split}
\Psi & = \frac{( r^1 + i r^2 )}{\sqrt{2}}
           + \sqt \tp \chi_+ + \sqt \tm \chi_- + 2 \tp \tm G
	   + \dotsb\\
\Theta & = \frac{( r^3 - i \theta )}{\sqrt{2}}
           - i \sqt \tp \bar{\tilde{\chi}}_+
           - i \sqt \tmb \tilde{\chi}_- + 2 \tp \tmb \tG
	   + \dotsb \quad.
\end{split}
\end{equation}
This twisted hypermultiplet is exchanged under T-duality with the hypermultiplet of the Kaluza-Klein monopole; its two chiral multiplets $\Psi'$ and $\Gamma$ are:
\begin{equation}
\label{eq:gammamult}
\begin{split}
\Psi' & = \frac{( {r^1}' + i {r^2}' )}{\sqrt{2}}
           + \sqt \tp \chi_+' + \sqt \tm \chi_-' + 2 \tp \tm G'
	   + \dotsb\\
g^2 \Gamma & = \frac{( {r^3}' + i g^2 \gamma )}{\sqrt{2}}
           - i \sqt \tp \bar{\tilde{\chi}}_+'
           - i \sqt \tm \bar{\tilde{\chi}}_-' + 2 \tp \tm \tG'
	   + \dotsb \quad.
\end{split}
\end{equation}
The primed fields in Eq.~\eqref{eq:gammamult} are exchanged with the corresponding unprimed fields in Eq.~\eqref{eq:twistedmult} under T-duality (except for $\tG'$). Generalizing the bosonic case of Eq.~\eqref{eq:KKGLSMbos}, gauge invariance of the action will require $\Gamma$ to transform by a simple shift: $\Gamma \to \Gamma + i \sqt \Lambda$.

\bigskip

The supersymmetric gauged linear sigma model action for the NS5-brane
is the sum of the following Lagrangian terms (plus complex conjugates
of the $F$ and $\widetilde{F}$ terms):
\begin{equation}
\label{eq:NS5GLSMsusy}
\begin{gathered}
\mathcal{L}_{\text{NS5},D} = \int d^4\theta \left[
  \frac{1}{e^2} \left( -\Sigma^\dag \Sigma + \Phi^\dag \Phi \right)
  + \frac{1}{g^2} \left( -\Theta^\dag \Theta + \Psi^\dag \Psi \right)
  + Q^\dag e^{2 V} Q + \Qt^\dag e^{-2 V} \Qt \right]\\
\mathcal{L}_{\text{NS5},F} = \int d^2\theta \,
  \left( \sqt \Qt \Phi Q - \Phi \Psi \right) \hspace{5em}
\mathcal{L}_{\text{NS5},\widetilde{F}} = \int d^2\vartheta \left( 
-\Theta \Sigma \right) \quad.
\end{gathered}
\end{equation}
Here, $d^2\theta \equiv -d\tp d\tm/2$ and $d^2\vartheta \equiv -d\tp d\tmb/2$ are the usual measures on chiral and twisted chiral superspace. The $\widetilde{F}$ term and its conjugate can be rewritten as an integral over full superspace:
\begin{equation}
\label{eq:Ftconvert}
\calL_\text{$\Theta$--$V$} =
  \int d^2\vartheta \bigl( -\Theta \Sigma \bigr)
  + \int d^2\bar{\vartheta} \bigl( -\Theta^\dag \Sigma^\dag \bigr)
  = \int d^4\theta \left[
    -\sqt \left( \Theta + \Theta^\dag \right) V \right]
    - \euab \partial_a (\theta A_b) \quad.
\end{equation}
The final term is the same topologically significant total derivative that appeared in Eq.~\eqref{eq:KKGLSMbos}.

After integrating out auxiliary fields, the component expansion of Eq.~\eqref{eq:NS5GLSMsusy} has bosonic part identical to Eq.~\eqref{eq:NS5GLSMbos} from the previous section plus terms involving the vector multiplet scalars. In the low energy limit those scalars are auxiliary fields and are integrated out with the rest of the vector multiplet. The full component expansion including fermions is given in~\cite{Harvey:2005ab}; it is very similar to the expansion given below for the doubled case, so we will not repeat it here.

The component action at low energy can be understood geometrically in terms of the real scalars $r^\mu = \{r^1, r^2, r^3, \theta\}$ and their real superpartners $\Omega^\mu$. The general supersymmetric nonlinear sigma model Lagrangian is~\cite{Howe:1985pm}:
\begin{multline}
\label{eq:nonlinsigmacompts}
\calL = - \frac{1}{2} \eta^{ab} G_{\mu \nu} \partial_a
  r^\mu \partial_b r^\nu 
  - \frac{1}{2} \euab B_{\mu \nu} \partial_a r^\mu \partial_b
  r^\nu \\
  \mbox{}
  + \frac{i}{2}  g_{\mu \nu} \Omega_-^\mu D_+ \Omega_-^\nu
  + \frac{i}{2}  g_{\mu \nu} \Omega_+^\mu D_- \Omega_+^\nu
  + \frac{1}{4} R_{\mu \nu \rho \lambda}
      \Omega_+^\mu \Omega_+^\nu \Omega_-^\rho \Omega_-^\lambda
  \quad,
\end{multline}
where in this case $G_{\mu \nu}$ and $B_{\mu \nu}$ correspond to the smeared NS5-brane solution. In the conventions used in this equation (only), $D_\pm \equiv D^\pm_0 \pm D^\pm_1$, where $D^\pm_a$ is the covariant derivative defined with positive or negative torsion, respectively. $R_{\mu \nu \rho \lambda}$ is the Riemann tensor defined with positive torsion.

\bigskip

As derived in~\cite{Harvey:2005ab} (via a supersymmetric generalization of the first order form method in section~\ref{sec:bosGLSMconv}), the T-dual supersymmetric gauged linear sigma model action for the Kaluza-Klein monopole is the sum of the following Lagrangian terms:
\begin{equation}
\begin{gathered}
\mathcal{L}_{\text{KK},D}  = \int d^4\theta \biggl[
  \frac{1}{e^2} \left( -\Sigma^\dag \Sigma + \Phi^\dag \Phi \right)
  + \frac{g^2}{2}
      \left( \Gamma + \Gamma^\dag - \sqt V \right)^2
\hspace{10em} \\
\hspace{16em} \mbox{}
 + \frac{1}{g^2} {\Psi'}^\dag \Psi'
  + Q^\dag e^{2 V} Q + \Qt^\dag e^{-2 V} \Qt \biggr] \\
\label{eq:KKGLSMsusy}
\mathcal{L}_{\text{KK},F}  = \int d^2\theta \, 
  \left( \sqt \Qt \Phi Q - \Phi \Psi' \right) \hspace{5em}
\calL_{\text{KK},\text{top}} = \euab \partial_a \left[\theta
  (\partial_b \gamma - A_b) \right]
 \quad,
\end{gathered}
\end{equation}
The topological term involving $\theta$ also appeared in the bosonic case; here, its origin can be traced to Eq.~\eqref{eq:Ftconvert}.

As in the NS5-brane case, the component form of this action agrees with the bosonic terms in Eq.~\eqref{eq:KKGLSMbos} plus vector multiplet scalar terms. When interpreting the low energy limit, it is important to write the action entirely in terms of the gauge invariant scalar $r^4 = \tht \equiv \gamma + \alpha$ and its gauge invariant superpartner $\Omega^4$: the superpartner of the component scalar $\gamma$ has more complicated supersymmetry transformations because Wess-Zumino gauge is not naturally preserved for gauge-variant objects. The end result is again of the form in Eq.~\eqref{eq:nonlinsigmacompts} with the conventional background for the KK-monopole.

\subsubsection{A doubled supersymmetric GLSM for monopoles}
\label{sec:doubledGLSMsusy}

The doubled bosonic GLSM discussed in section~\ref{sec:GLSMbos} has a
natural extension to the $\calN = (4,4)$ case considered here. However, simply doubling the fiber coordinate would not preserve supersymmetry: the full (4,4) (twisted) hypermultiplet must be doubled, including components corresponding to coordinates on the base. This is reminiscent of the ``doubled everything'' discussion in~\cite{Hull:2006va}. As in the previous examples, the appropriate doubled Lagrangian is the mean of those corresponding to the two choices of polarization:
\begin{equation}
\label{eq:doubledGLSMsusy}
\begin{gathered}
\begin{aligned}
\mathcal{L}_{\text{d},D}  = \int d^4\theta \biggl[
  &\frac{1}{e^2} \left( -\Sigma^\dag \Sigma + \Phi^\dag \Phi \right)
  + \frac{1}{2g^2} \left( -\Theta^\dag \Theta + \Psi^\dag \Psi \right)
  - \frac{\sqt}{2} \left( \Theta + \Theta^\dag \right) V
\\
  & \mbox{} + \frac{g^2}{4} \left( \Gamma + \Gamma^\dag - \sqt V
  \right)^2
  + \frac{1}{2 g^2} {\Psi'}^\dag \Psi'
  + Q^\dag e^{2 V} Q + \Qt^\dag e^{-2 V} \Qt
   \biggr]
\end{aligned} \quad~\\
\mathcal{L}_{\text{d},F}  = \int d^2\theta \, 
  \left[ \sqt \Qt \Phi Q - \frac{1}{2} \Phi (\Psi + \Psi') \right]
\qquad\qquad
\calL_{\text{d},\text{top}} = \frac{1}{2} \euab \partial_a \left[\theta
  (\partial_b \gamma - 2 A_b) \right]
 \quad.
\end{gathered}
\end{equation}
Here, the $\calL_{\widetilde{F}}$ terms from the NS5-brane GLSM have been rewritten as $\calL_\text{$\Theta$--$V$}$ from Eq.~\eqref{eq:Ftconvert}.\footnote{As in Eq.~\eqref{eq:Ftconvert}, gauge invariance is not manifest after this replacement. This could be naturally addressed in the doubled formalism by adding and subtracting a term mixing $\Theta$ and $\Gamma$, as superfields and in components: $\int\! d^4\theta \left[ -\frac{\sqrt{2}}{2} \left( \Theta + \Theta^\dag \right) V \right] - \frac{1}{2} \euab \partial_a (\theta A_b) = 
 \int\! d^4\theta \left[ \frac{1}{2} \left( \Theta + \Theta^\dag \right) \left( \Gamma + \Gamma^\dag - \sqt V \right) \right] + \frac{1}{2} \euab \partial_a \left[ \theta   (\partial_b \gamma - A_b) \right]$.}  Just as we found in the bosonic case, this already includes the topological term from the doubled formalism. This system inherits full $\calN=(4,4)$ supersymmetry from the NS5-brane and KK-monopole GLSMs since it is simply a sum of the two.

After integrating out the auxiliary fields, the component action can be written as the sum of kinetic, scalar potential, and ``Yukawa'' parts, plus a mixing term left over from $\calL_{d,\text{top}}$: $\calL = \calL_\text{kin} + \calL_\text{pot} + \calL_\text{Yuk} + \calL_\text{mix}$. The kinetic term reads:
\begin{multline}
\label{eq:LkinCompts}
\mathcal{L}_{\text{kin}} =
     \frac{1}{e^2} \Bigl( \frac{1}{2} F^{2}_{01} 
       - |\partial_a \phi|^2 - |\partial_a \sigma|^2
       + i (\bar{\lambda}_+ \partial_- \lambda_+ +
            \bar{\tilde{\lambda}}_+ \partial_- \tilde{\lambda}_+ +
	    \bar{\lambda}_- \partial_+ \lambda_- +
            \bar{\tilde{\lambda}}_- \partial_+ \tilde{\lambda}_- )
       \Bigr) \quad \; \\
\shoveleft{ \qquad \quad
 \mbox{} + \frac{1}{2g^2} \Bigl( -\frac{1}{2} (\partial_a r^m)^2
       -\frac{1}{2} (\partial_a {r^m}')^2
       - \frac{1}{2} (\partial_a \theta)^2
       - \frac{g^4}{2} (\partial_a \gamma - A_a)^2 }\\
  \shoveright{\mbox{}
       + i (\bar{\chi}_+ \partial_- \chi_+ +
            \bar{\tilde{\chi}}_+ \partial_- \tilde{\chi}_+ +
	    \bar{\chi}_- \partial_+ \chi_- +
            \bar{\tilde{\chi}}_- \partial_+ \tilde{\chi}_- )
       \hspace*{0.8cm} } \\
  \shoveright{\mbox{}
       + i (\bar{\chi}_+' \partial_- \chi_+' +
            \bar{\tilde{\chi}}_+' \partial_- \tilde{\chi}_+' +
	    \bar{\chi}_-' \partial_+ \chi_-' +
            \bar{\tilde{\chi}}_-' \partial_+ \tilde{\chi}_-' )
       \Bigr) \quad \;} \\
\shoveleft{ \qquad \quad
 \mbox{} + \Bigl( 
       - |\calD_a q|^2 - |\calD_a \qt|^2 }
       + i (\bar{\psi}_+ \calD_- \psi_+ +
            \bar{\tilde{\psi}}_+ \calD_- \tilde{\psi}_+ +
	    \bar{\psi}_- \calD_+ \psi_- +
            \bar{\tilde{\psi}}_- \calD_+ \tilde{\psi}_- )
       \Bigr) \quad.
\end{multline}
In the fermion kinetic terms, $\partial_\pm \equiv \partial_0 \pm \partial_1$; the covariant derivatives of the charged hypermultiplet components were defined below Eq.~\eqref{eq:NS5GLSMbos}. It is helpful to define average and relative combinations of the doubled components:
\begin{equation}
\begin{aligned}
\label{eq:ravgdef}
\ra^m &\equiv \frac{1}{2} (r^m + {r^m}') \;, &\quad&
&\rrel^m &\equiv \frac{1}{2} (r^m - {r^m}') \;, \\
\chi_{\text{avg},\pm} &\equiv \frac{1}{2} (\chi_\pm + \chi_\pm') \;, 
 &&
&\chi_{\text{rel},\pm} &\equiv \frac{1}{2} (\chi_\pm - \chi_\pm') \;,
\end{aligned}
\end{equation}
and similarly for $\tilde{\chi}_\text{avg}$ and $\tilde{\chi}_\text{rel}$. The $r^m$ kinetic terms then take the form
\begin{equation}
\label{eq:rkinetic}
-\frac{1}{4g^2} (\partial_a
r^m)^2 -\frac{1}{4g^2} (\partial_a {r^m}')^2 = - \frac{1}{2g^2}
(\partial_a \ra^m)^2 - \frac{1}{2g^2} (\partial_a \rrel^m)^2 \quad.
\end{equation}
The $\chi$--$\chi'$ fermion kinetic terms split into average and relative parts in the same way.

After integrating out the auxiliary fields, the scalar potential  reads:
\begin{equation}
\label{eq:LpotCompts}
\begin{split}
\mathcal{L}_{\text{pot}} & =
     - \frac{e^2}{2} \left( |q|^2 - |\qt|^2 - \ra^3 \right)^2
     - \frac{e^2}{2} \left| 2 \qt q 
       - \left( \ra^1 + i \ra^2 \right) \right|^2 \\
 & \quad \, \mbox{} - \left( |\phi|^2 + |\sigma|^2 \right)
       \left( g^2 + 2 |q|^2 + 2 |\qt|^2 \right) 
   + \theta F_{01}
 \quad.
\end{split}
\end{equation}
The relative coordinates $\rrel^m$ decouple in all interactions. The final term of the equation above includes contributions from $\calL_{d,\text{top}}$; the remainder of the topological term gives a cross term mixing the doubled coordinates:
\begin{equation}
\label{eq:Lmix}
\calL_\text{mix} = \frac{1}{2} \euab \partial_a \theta
 (\partial_b \gamma - A_b) \quad.
\end{equation}

Finally, the fermion interactions are contained in the ``Yukawa'' term (which also includes several two-fermion interactions):
\begin{equation}
\label{eq:LYukCompts}
\begin{split}
\mathcal{L}_{\text{Yuk}} & = \;\;\;
     \left( \tilde{\lambda}_+ \chi_{\text{avg},-}
     - \lambda_+ \bar{\tilde{\chi}}_{\text{avg},-}
     + \bar{\lambda}_+ \tilde{\chi}_{\text{avg},- }
     - \bar{\tilde{\lambda}}_+ \bar{\chi}_{\text{avg},-} \right)\\
 & \quad \mbox{}
   - \left( \tilde{\lambda}_- \chi_{\text{avg},+ }
     - \lambda_- \bar{\tilde{\chi}}_{\text{avg},+}
     + \bar{\lambda}_- \tilde{\chi}_{\text{avg},+}
     - \bar{\tilde{\lambda}}_- \bar{\chi}_{\text{avg},+} \right)\\
 & \quad \mbox{} + i \sqt q \; \left(
       \bar{\lambda}_+ \bar{\psi}_- - \bar{\lambda}_- \bar{\psi}_+
       + i \tilde{\lambda}_+ \tilde{\psi}_- 
       - i \tilde{\lambda}_- \tilde{\psi}_+ \right)
     + \sqt \, \sigma \; \left( \psi_+ \bar{\psi}_-
       - \tilde{\psi}_+ \bar{\tilde{\psi}}_- \right)\\
 & \quad \mbox{} + i \sqt q^\dag \left(
       \lambda_+ \psi_- - \lambda_- \psi_+
       - i \bar{\tilde{\lambda}}_+ \bar{\tilde{\psi}}_- 
       + i \bar{\tilde{\lambda}}_- \bar{\tilde{\psi}}_+ \right)
     - \sqt \sigma^\dag \left( \bar{\psi}_+ \psi_-
       - \bar{\tilde{\psi}}_+ \tilde{\psi}_- \right)\\
 & \quad \mbox{} - i \sqt \qt \; \left(
       \bar{\lambda}_+ \bar{\tilde{\psi}}_- 
       - \bar{\lambda}_- \bar{\tilde{\psi}}_+
       - i \tilde{\lambda}_+ \psi_- 
       + i \tilde{\lambda}_- \psi_+ \right)
     - \sqt \, \phi \; \left( \psi_+ \tilde{\psi}_-
       + \tilde{\psi}_+ \psi_- \right)\\
 & \quad \mbox{} - i \sqt \qt^\dag \left(
       \lambda_+ \tilde{\psi}_- - \lambda_- \tilde{\psi}_+
       + i \bar{\tilde{\lambda}}_+ \bar{\psi}_- 
       - i \bar{\tilde{\lambda}}_- \bar{\psi}_+ \right)
     + \sqt \phi^\dag \left( \bar{\psi}_+ \bar{\tilde{\psi}}_-
       + \bar{\tilde{\psi}}_+ \bar{\psi}_- \right) \quad.
\end{split}
\end{equation}
As in the scalar potential, only the $\chi_\text{avg}$ fields appear in interactions: the $\chi_\text{rel}$ fields decouple.

The bosonic part of this complete action reduces precisely to the doubled bosonic GLSM of Eq.~\eqref{eq:doubledGLSMbos} after replacing $\ra^m \to r^m$ and dropping the (decoupled) $\rrel^m$ kinetic term and terms involving the vector multiplet scalars. This action as a whole is nearly identical to the component actions for the NS5-brane and KK-monopole; apart from the decoupled relative kinetic terms (and $\ra^m \to r^m$), only the terms involving $\theta$ or $\gamma$ differ between the three.

\bigskip

To relate this to the GLSM for the NS5-brane by imposing a constraint between the doubled fields, we can generalize the procedure of sections~\ref{sec:DoubledSmearedConstraint} and~\ref{sec:bosGLSMdoub} by adding a (non-dynamical) $\calN=(4,4)$ vector multiplet to gauge away the dynamics of $\gamma$ and the other components of the $\Psi'$--$\Gamma$ hypermultiplet. At the $\calN=(2,2)$ superfield level, the gauge terms added to the overall Lagrangian depend on an auxiliary vector superfield $B$ and its accompanying chiral superfield $P$ (and conjugate $P^\dag$):
\begin{equation}
\label{eq:SUSYgauged}
\begin{aligned}
\calL_{g,D} &= \int d^4\theta \left[ \frac{1}{4g^2} B^2
  + \frac{1}{2 g^2} B \left( \Theta + \Theta^\dag
    - g^2 (\Gamma + \Gamma^\dag - \sqt V) \right) \right]\\
\calL_{g,F} &= \int d^2\theta \left[ \frac{\sqt}{4g^2} P
  \left( \Psi - \Psi' \right) \right] \quad.
\end{aligned}
\end{equation}
The superfield $P$ acts as a Lagrange multiplier, forcing $\Psi = \Psi'$; at the component level this equates all the components of $\Psi$ with the corresponding primed components of $\Psi'$. Supersymmetric gauge transformations of $B$ absorb any difference between the components of $\Theta$ and the corresponding primed components of $\Gamma$. Finally, the vector field component of $B$ acts precisely as the gauge field $C_a$ in section~\ref{sec:DoubledSmearedConstraint}: its gauge transformations absorb the dynamics of $\gamma$.

This is explicit at the superfield level with a field redefinition reminiscent of the one in that earlier section. Defining a vector superfield $D \equiv B + \Theta + \Theta^\dag - g^2 (\Gamma + \Gamma^\dag - \sqt V)$, its square absorbs the gauge terms and many terms from the action:
\begin{equation}
\label{eq:superauxterm}
\begin{split}
  \frac{1}{4g^2} D^2 = &\frac{1}{4g^2} B^2 + \frac{1}{2g^2} B \left(
    \Theta + \Theta^\dag - g^2 (\Gamma + \Gamma^\dag - \sqt V)
  \right) + \frac{1}{4g^2} (\Theta + \Theta^\dag)^2\\
  &\mbox{} + \frac{\sqrt{2}}{2} (\Theta + \Theta^\dag) V +
  \frac{g^2}{4} (\Gamma + \Gamma^\dag - \sqt V)^2 - \frac{1}{2}
  (\Theta + \Theta^\dag) (\Gamma + \Gamma^\dag) \quad.
\end{split}
\end{equation}
Then up to unimportant total derivative terms, the $D$ terms can be written as $\calL_{\text{d},D}+\calL_{g,D} = \calL_{\text{NS5},D}+\calL_\text{$\Theta$--$V$}+\int d^4\theta [\frac{1}{4g^2} D^2]$, and the doubled $F$ term reduces to the NS5-brane F-term due to the $P$ equation of motion. Thus, the doubled GLSM plus this choice of gauge terms is equivalent to the NS5-brane GLSM of Eq.~\eqref{eq:NS5GLSMsusy}, plus the decoupled auxiliary superfield.

\bigskip

The steps to relate the doubled GLSM to that of the KK-monopole by gauging away the dynamics of $\theta$ and the other $\Psi$--$\Theta$ components are remarkably similar. Formally, the only change is to subtract the terms of Eq.~\eqref{eq:SUSYgauged} from the doubled action rather than adding them. Then $\calL_{\text{d},D}-\calL_{g,D} = \calL_{\text{KK},D}+\calL_{\text{KK},\text{top}}-\int d^4\theta [\frac{1}{4g^2} D^2]$, and the doubled $F$ term reduces appropriately as well: the doubled GLSM with these gauge terms is equivalent to the KK-monopole GLSM of Eq.~\eqref{eq:KKGLSMsusy} plus the decoupled auxiliary superfield.

The one subtlety in this case is that for the vector superfield $B$ to correctly gauge away the dynamics of $\theta$ and the other components of the twisted chiral superfield $\Theta$, its gauge transformations $B \to B + i(\Lambda - \Lambda^\dag)$ must take $\Lambda$ to likewise be a twisted chiral superfield rather than a chiral one as usual. This in turn implies minor changes to the interpretation and relationships of some vector multiplet components. It would be interesting to understand how such an object's behavior would differ from that of a usual vector multiplet. But in this case, $B$ appears only as a non-dynamical auxiliary field and this non-standard gauging has no effect on the remainder of the calculation.

\subsection{Instanton corrections and localization}
\label{sec:localization}

Although the calculations above seem at first to show that these gauged linear sigma models describe only the smeared versions of the NS5-brane and KK-monopole, this understanding is incomplete. For both the NS5-brane~\cite{Tong:2002rq} and the KK-monopole~\cite{Harvey:2005ab}, a sum over worldsheet instantons in the GLSM yields corrections to the low energy effective action. These corrections are expected in multiple terms of the action, but here we consider only their effect on the four-fermion term. As can be seen from Eq.~\eqref{eq:nonlinsigmacompts}, corrections to this term can be interpreted as modifying the target space curvature in the low energy limit.

Surprisingly, the relevant terms in the doubled GLSM are precisely the same as in those earlier calculations, meaning that the instanton calculation and its results are also formally identical (though of course the interpretation must be adjusted to fit this new context). In fact, in the doubled formalism the calculation clarifies an ambiguity in the KK-monopole calculation of~\cite{Harvey:2005ab}, suggesting once again that the doubled picture is the proper context for understanding this background. Because of the similarity between this calculation and the previous work in the references, the discussion below will simply summarize the key computational steps while making note of subtleties associated with this doubled context.

\subsubsection{Setting up the instanton calculation}
\label{sec:instantonSetup}

In a gauged linear sigma model, worldsheet instantons correspond to vortices of the gauge field, counted by a line integral of $A_a$ at infinity or equivalently by
\begin{equation}
\label{eq:instantonCount}
k = -\frac{1}{2 \pi} \int F_{1 2} \quad.
\end{equation}
The target space of this GLSM (with one unit of monopole charge) does not contain any holomorphic two-cycles, so no worldsheet instantons from holomorphic embeddings of the worldsheet are present. (The $T^2$ parametrized by $\theta$ and $\gamma$ does not provide such an embedding because it cannot be viewed as a complex manifold: the constraint from the doubled formalism cuts the degrees of freedom of its coordinates in half.)  Instead, the instantons appearing here correspond to the ``point-like instantons'' of~\cite{Witten:1993yc}: they are smooth solutions of the underlying GLSM but appear singular in the effective low energy theory.

The first step toward finding the instanton sum is to compute the bosonic action in each instanton sector $k$, based on the bosonic terms in the doubled GLSM action from Eqs.~\eqref{eq:LkinCompts}--\eqref{eq:Lmix}. This calculation begins with the choice of a classical solution to represent each instanton sector. Strictly speaking, this approach is not valid in 1+1 dimensions: the lack of symmetry breaking implied by the Mermin-Wagner-Coleman theorem~\cite{Mermin:1966fe,Coleman:1973ci} means that the language of classical vacua and moduli spaces does not apply. In particular, there are no finite action solutions of the equations of motion in this system that satisfy the vortex boundary conditions.

To deal with this difficulty, we apply a version of the ``constrained instantons'' method~\cite{Affleck:1980mp}. The calculation is performed in the limit of vanishing Taub-NUT radius ($g \to 0$) where BPS solutions exist, and supersymmetry protects the result when that limit is relaxed. Because the instanton corrections are finite even in the strict $g \to 0$ limit, this procedure gives a meaningful contribution to the low energy action.

Working in that framework, we begin as planned by choosing a specific classical vacuum for the instanton solutions to approach at large distance. The potential terms in the action imply that $\phi = \sigma = 0$ in all vacua. The R-symmetry of the action allows us to specialize to the case $\qt=0$ without loss of generality, at which point the potential also requires $r^1_\text{avg} = r^2_\text{avg} = 0$ and $|q|^2=r^3_\text{avg}=\zeta$ (where $\zeta$ is a constant parametrizing the vacuum).

Only the independent bosonic variations about this chosen vacuum that could affect the gauge field $A_a$ are significant for the instanton calculation, so we truncate to that set. After a Wick rotation to Euclidean space (taking $\sigma^2 \equiv i \sigma^0$) the truncated action reads
\begin{multline}
\label{eq:EucBosAction}
S_{\text{E}} = \frac{i}{2 \pi} \int d^2\sigma \left[
  \frac{1}{2 e^2} F_{12}^2
  + \frac{1}{2 g^2} \left( \partial_a \ra^3 \right)^2
  + \frac{1}{4g^2}  \left( \partial_a \theta \right)^2
  + \frac{g^2}{4}   \left( \partial_a \gamma - A_a \right)^2
  \right. \\ \left. \mbox{}
  - \frac{1}{2} \euab \partial_a \theta (\partial_b \gamma - A_b)
  + \left| \calD_a q \right|^2
  + \frac{e^2}{2} \left( |q|^2 - r_\text{avg}^3 \right)^2 
  + i \theta F_{12}
  \right] \quad. 
\end{multline}
When $g \to 0$, the $\gamma$ kinetic term drops out entirely, while variations in $\theta$ and $\ra^3$ are completely frozen out (even when variations in $q$ make $|q|^2 \ne \zeta$). Thus, all terms involving derivatives of $\theta$ drop from the action: the doubled formalism explicitly justifies the assumption in~\cite{Harvey:2005ab} that $\theta$ can be treated as be constant. Thus, the relevant terms in the instanton action are
\begin{equation}
\label{eq:instantonAction}
S = \frac{i}{2 \pi} \int d^2 \sigma \left[
  \frac{1}{2 e^2} F_{12}^2
  + \left| \calD_\mu q \right|^2
  + \frac{e^2}{2} \left( |q|^2 - \zeta \right)^2 
  + i \theta F_{12}
  \right] \quad.
\end{equation}
This is identical to the bosonic action found previously for both the NS5-brane and KK-monopole: the abelian Higgs model action at critical coupling plus a $\theta$ term. The fermionic terms relevant in this limit are identical to those in previous treatments as well.

Because the actions relevant to the instanton calculation are precisely the same in all three cases, those earlier calculations follow through immediately here. Completing the square in the bosonic action and minimizing the action in each instanton sector yields a set of first order Bogomol'nyi equations for the vortex solutions:
\begin{equation}
\label{eq:Bogomolnyi}
F_{12} = \pm e^2 (|q|^2 - \zeta)
 \qquad \text{and} \qquad
\mathcal{D}_1 q \pm i \mathcal{D}_2 q = 0 \quad.
\end{equation}
When these equations are satisfied, the action reduces to $S = \frac{i}{2\pi} \int d^2\sigma (\mp \zeta + i \theta) F_{12}$, which integrates to give constants times the instanton number $k$. After selecting the $\pm$ sign to give the tightest bound on the real part of the action (depending on the sign of $k$), the resulting path integral factor $e^{iS}$ in each instanton sector $k$ can be written as $e^{-S_k}$, with
\begin{equation}
\label{eq:VortexAction}
S_k = |k| \zeta - i k \theta \quad.
\end{equation}

\subsubsection{Corrections to the four-fermion term}
\label{sec:fourFermiCorrections}

A complete sum over instanton configurations consists of a discrete sum over sectors $k$ and an integral over the zero modes within each sector. The references provide the details of this calculation, which are entirely unchanged here: identifying bosonic and fermionic zero modes to find the proper measure for the integral, finding how the instanton action is modified due to interference between fermion zero modes, and determining the asymptotic behavior of the zero mode solutions themselves.

As in the previous work on the NS5-brane and KK-monopole systems individually, the most straightforward way to understand the impact of the instanton sum on the geometry is to compute its effect on the effective four-fermion interaction at low energy. The $k$-instanton contribution to the $\psi^4$ correlation function for $k>0$ is $G_4^{(k)}(\sigma_1,\sigma_2,\sigma_3,\sigma_4) = \bigl. \bigl< \bar{\psi}_+(\sigma_1) \psi_-(\sigma_2) \tilde{\psi}_+(\sigma_3) \bar{\tilde{\psi}}_-(\sigma_4) \bigr> \bigr|_{k\text{-instanton}}$ (or for $k<0$, the conjugate). The total modification to the four-fermion term in the low energy action is then $-\sum_k G_4^{(k)}$ (where the minus sign appears after Wick rotation back to the Lorentzian worldsheet).

As found in~\cite{Tong:2002rq,Harvey:2005ab}, the result of this calculation is an additional $\psi^4$ contribution to the low energy effective action:
\begin{equation}
\label{eq:LeffCorrection}
\delta \mathcal{L}_{\text{eff}, \psi^4} =
  - \sum_{k=1}^\infty \frac{2 r}{\pi k}
  \nu(\tilde{\mathcal{M}}_k) e^{-k r}
  \left[
  e^{i k \theta}
    \bar{\psi}_+ \psi_- \tilde{\psi}_+ \bar{\tilde{\psi}}_-
  + e^{-i k \theta}
    \bar{\psi}_- \psi_+ \tilde{\psi}_- \bar{\tilde{\psi}}_+
  \right] \quad,
\end{equation}
where $r$ has been restored in place of the vacuum parameter $\zeta$, and $\nu(\tilde{\mathcal{M}}_k)$ is a specific integral over the moduli space of the relative positions of $k$ vortices whose value can be determined by comparison to the known supergravity result for the localized NS5-brane~\cite{Tong:2002rq}.

Those references show that for the individual monopole actions, this additional $\psi^4$ contribution is related in the low energy limit to a $\chi^4$ term, which can in turn be related to a sum of terms quadratic in the real fermions $\Omega^\mu_\pm$ that appear in the nonlinear sigma model action of Eq.~\eqref{eq:nonlinsigmacompts}. As previously discussed, for the NS5-brane case these terms change the $(\Omega)^4$ terms in that action to match the Riemann tensor of the known localized NS5-brane background at large $r$ (and keeping only leading order terms in $g$). For the KK-monopole case, the changes to the effective Riemann tensor are consistent with the localized metric given in Eq.~\eqref{eq:locKKmetric} and the torsion that immediately follows it.

The doubled GLSM formalism presented here relates all these results in a natural way. The contribution to the effective $\psi^4$ interaction from worldsheet instantons is present in the doubled picture, but it does not initially have clear geometric significance: its impact on the target space Riemann tensor can only be determined after a choice of polarization allows a geometric interpretation of the action. Once a polarization is chosen, the low energy background fields (including the instanton corrections depending on $\theta$) are related by the standard Buscher rules as the doubled formalism requires.

\section{Conclusions}
\label{sec:Conclusions}

\subsection{Overview of results}
\label{sec:Results}

The Kaluza-Klein monopole/NS5-brane system provides a valuable example of the doubled geometry formalism for symmetric treatment of T-dual string backgrounds. In its smeared approximation this system can act as a simple but nontrivial illustration of the formalism, but the true localized solution is more important: it is an example of a familiar system that cannot be fully described in a conventional formulation. The need in this case for a formalism based on a symmetric treatment of the geometric coordinate and its non-geometric dual was explicitly recognized by Harvey and the present author in~\cite{Harvey:2005ab}, where the form of the localized KK-monopole background was first partially computed. That work explicitly asserted that the KK-monopole's ``winding space coordinate'' $\theta$ (the Fourier dual to string winding charge) could only contribute the necessary features to the solution if it were a dynamical worldsheet field. The incompatibility of this assertion with a conventional formalism may have made it natural to doubt that conclusion~\cite{Okuyama:2005gx}, but the success of the present doubled geometry analysis confirms that this is the correct description of the system.

Despite the broken $S^1$ isometry of the NS5-brane background, the doubled picture appears to be a complete and consistent description of both sides of the T-dual pair. While it is perhaps surprising to find that the Buscher rules apply even in this case, they arise very generally in the doubled formalism and their result is entirely consistent with the earlier calculation of worldsheet instanton corrections in~\cite{Harvey:2005ab}. The geometrical insight provided by doubled geometry may prove to be a more fruitful way of understanding this and similar systems than was possible when~\cite{Gregory:1997te} described the effects simply as a coherent state of classical string winding modes in string field theory or in earlier treatments of winding mode expansions in T-duality such as~\cite{Shapere:1988zv,Giveon:1988tt}. It would be interesting to make the relationship between these descriptions more precise.

Extending the doubled formalism from the usual nonlinear sigma models which have a direct geometrical interpretation to gauged linear sigma models which do not broadens its potential applications. This mathematical structure is consistent at least at the classical level in both the bosonic and supersymmetric models. The success of the worldsheet instanton calculation when starting from the doubled GLSM action gives good reason to believe that this method is valid. As noted previously, it provides the missing justification for assuming $\theta$ is constant in the instanton calculation of~\cite{Harvey:2005ab}, and of course doubled geometry is required for a sensible interpretation of that calculation's results.

\subsection{Interpretation and future directions}
\label{sec:Comments}

The perspective of this work, as in~\cite{Tong:2002rq,Harvey:2005ab}, is that the localized versions of the NS5-brane and Kaluza-Klein monopole are the ``true'' descriptions of those objects in string theory: only they show the expected ``throat'' behavior near the monopole cores, and only they can be properly associated with 2D QFTs. If one attempts to formulate string theory on the smeared backgrounds, worldsheet instantons are sensitive to the physics that requires this localization and their contributions will fill in the corresponding low energy effects.

It is interesting to ask how the Kaluza-Klein dyon collective coordinate relates to this doubled picture. As considered by Sen~\cite{Sen:1997zb} and by Gregory, Harvey, and Moore~\cite{Gregory:1997te}, the dyon coordinate is a zero-energy gauge deformation of the standard Taub-NUT supergravity background. Taken as a constant parameter, its only contribution to the action is an overall phase, although a time-dependent value does allow the monopole to carry Kaluza-Klein electric charge. As shown in~\cite{Harvey:2005ab}, the Buscher rules map the KK-dyon coordinate to a shift in the (smeared) NS5-brane's $\theta$ coordinate. In the doubled picture, the field $\theta$ in the KK-monopole solution naturally plays precisely the role of the dyon parameter (as discussed at the end of section~\ref{sec:bosGLSMconv}): constant shifts in $\theta$ have zero energy and its momentum carries KK electric charge. The difference is that here, the significance of the classical background of string winding modes is explicit as part of the doubled geometry. Thus, the manner in which shifts in $\theta$ lead to physically distinct configurations is made clear.

\bigskip

A natural concern when considering corrections to a well-known string background like the KK-monopole is why they have not been observed in previous studies of the system. After all, the localized form of the KK-monopole metric in Eq.~\eqref{eq:locKKmetric} has a distinctly different $r$ dependence near the monopole core and its torsion is non-zero. Even if one considers only conventional geometry, shouldn't probes near that point be sensitive to these changes? How have they escaped detection?

To understand this, it is instructive to consider a specific example. In~\cite{Witten:2009xu}, Witten derives the standard Taub-NUT background metric as the T-dual of a (localized) NS5-brane by computing the moduli space of an intersecting brane configuration. Considering only the $\Reals^3 \times S^1$ of interest in the NS5-brane geometry, a D3-brane is taken to wrap $S^1$ but sits at a point in $\Reals^3$. T-duality leaves the moduli space unchanged but converts the D3-brane to a D2-brane localized on the dual $S^1$. Because the moduli space of a localized D-brane is simply its position, the relevant portion of the moduli space is precisely equal to the T-dual geometry, and the result of the calculation yields the Taub-NUT metric (with the dyonic $B$-field mode arising topologically). The geometry obtained appears to mach the ``smeared'' KK-monopole, with no trace of the corrections analyzed here.

The resolution of this puzzle is straightforward: a probe localized on the KK-monopole $S^1$ (like Witten's D2-brane) has by definition exactly zero string winding charge. Because winding charge is Fourier dual to the non-geometric coordinate $\theta$, the probe is therefore entirely delocalized in that direction and will be sensitive only to the (doubled) geometry as averaged over $\theta$. This averaging reproduces precisely the Taub-NUT metric and zero torsion. One would expect that for a D-brane instead \emph{wrapped} around this $S^1$, Wilson lines of the gauge field will act as coherent states of string winding modes with no definite winding number but corresponding to specific values of $\theta$. Thus, it is entirely reasonable that pointlike probes of the system would be sensitive only to the conventional geometry, averaging over any dependence on the non-geometric coordinate.

\bigskip

There are several natural extensions to this work. First, the evidence here suggests that the doubled formalism can be sensibly used to study T-duality even in the absence of an isometry, so there may be new insights to be gained by applying it to other known backgrounds localized on a circle. As in the present case, this would be unlikely to reveal any entirely unexpected physics, but it may bring greater clarity to those portions of the duality web. The biggest open question in these cases is quantization: for the formalism to be self-contained it would be helpful to be able to quantize the doubled theory without first choosing a polarization, and that ability will be absolutely essential for any treatment of generic string backgrounds with nontrivial dependence on both $\theta$ and $\tht$. The doubled GLSM methods employed here may also prove to be valuable for understanding a wider range of systems, but it would be highly desirable to have a rigorous mathematical understanding of the domain of validity of this approach.

Just as it provides a useful system for exploring the doubled geometry formalism in string theory, the KK-monopole/NS5-brane system may also prove to be an interesting background for the study of ``double field theory'' as developed in~\cite{Hull:2009mi,Hohm:2010jy,Hohm:2010pp} and related work. This field theory corresponds to the massless sector of closed string field theory with dependence on both momentum and winding modes, and it is defined on the same doubled torus introduced in doubled geometry. Thus, it can be viewed as the extension of Einstein gravity (coupled to a two-form and dilaton) to allow dependence on both position and ``winding space'' coordinates. This is precisely the right context in which to discuss the localized KK-monopole/NS5-brane background, which could provide a valuable example and test case in formulating double field theory for non-flat spaces.

\begin{acknowledgments}
I would like to thank Edward Witten for a helpful conversation regarding portions of the conclusions, the participants of the Great Lakes Strings 2011 for useful comments, and Alex Lundquist for productive discussions of duality and the KK-dyon mode. Portions of this work were completed at the Joint Science Department of Claremont McKenna, Pitzer, and Scripps Colleges; thanks to my colleagues there for hospitality and support.
\end{acknowledgments}

\bibliography{monopoles,nongeom}

\providecommand{\href}[2]{#2}\begingroup\raggedright\begin{thebibliography}{10}

\bibitem{Buscher:1987qj}
T.~H. Buscher, ``Path integral derivation of quantum duality in nonlinear sigma
  models,''
\href{http://dx.doi.org/10.1016/0370-2693(88)90602-8}{{\em Phys. Lett.}
  {\bfseries B201} (1988) 466}.

\bibitem{Rocek:1991ps}
M.~Ro{\v{c}}ek and E.~P. Verlinde, ``Duality, quotients, and currents,''
  \href{http://dx.doi.org/10.1016/0550-3213(92)90269-H}{{\em Nucl. Phys.}
  {\bfseries B373} (1992) 630--646},
\href{http://arxiv.org/abs/hep-th/9110053}{{\ttfamily arXiv:hep-th/9110053}}.

\bibitem{Gross:1983hb}
D.~J. Gross and M.~J. Perry, ``Magnetic monopoles in {K}aluza-{K}lein
  theories,''
\href{http://dx.doi.org/10.1016/0550-3213(83)90462-5}{{\em Nucl. Phys.}
  {\bfseries B226} (1983) 29}.

\bibitem{Sorkin:1983ns}
R.~D. Sorkin, ``{K}aluza-{K}lein monopole,''
\href{http://dx.doi.org/10.1103/PhysRevLett.51.87}{{\em Phys. Rev. Lett.}
  {\bfseries 51} (1983) 87--90}.

\bibitem{Banks:1988rj}
T.~Banks, M.~Dine, H.~Dykstra, and W.~Fischler, ``Magnetic monopole solutions
  of string theory,''
\href{http://dx.doi.org/10.1016/0370-2693(88)91233-6}{{\em Phys. Lett.}
  {\bfseries B212} (1988) 45}.

\bibitem{Ooguri:1995wj}
H.~Ooguri and C.~Vafa, ``Two-dimensional black hole and singularities of {CY}
  manifolds,'' \href{http://dx.doi.org/10.1016/0550-3213(96)00008-9}{{\em Nucl.
  Phys.} {\bfseries B463} (1996) 55--72},
\href{http://arxiv.org/abs/hep-th/9511164}{{\ttfamily arXiv:hep-th/9511164}}.

\bibitem{Gauntlett:1992nn}
J.~P. Gauntlett, J.~A. Harvey, and J.~T. Liu, ``Magnetic monopoles in string
  theory,'' \href{http://dx.doi.org/10.1016/0550-3213(93)90584-C}{{\em Nucl.
  Phys.} {\bfseries B409} (1993) 363--381},
\href{http://arxiv.org/abs/hep-th/9211056}{{\ttfamily arXiv:hep-th/9211056}}.

\bibitem{Khuri:1992ww}
R.~R. Khuri, ``A multimonopole solution in string theory,''
  \href{http://dx.doi.org/10.1016/0370-2693(92)91528-H}{{\em Phys. Lett.}
  {\bfseries B294} (1992) 325--330},
\href{http://arxiv.org/abs/hep-th/9205051}{{\ttfamily arXiv:hep-th/9205051}}.

\bibitem{Eyras:1998hn}
E.~Eyras, B.~Janssen, and Y.~Lozano, ``Five-branes, {KK}-monopoles and
  {T}-duality,'' \href{http://dx.doi.org/10.1016/S0550-3213(98)00575-6}{{\em
  Nucl.Phys.} {\bfseries B531} (1998) 275--301},
  \href{http://arxiv.org/abs/hep-th/9806169}{{\ttfamily arXiv:hep-th/9806169
  [hep-th]}}.

\bibitem{Gregory:1997te}
R.~Gregory, J.~A. Harvey, and G.~W. Moore, ``Unwinding strings and {T}-duality
  of {K}aluza-{K}lein and {H}-monopoles,'' {\em Adv. Theor. Math. Phys.}
  {\bfseries 1} (1997) 283--297,
\href{http://arxiv.org/abs/hep-th/9708086}{{\ttfamily arXiv:hep-th/9708086}}.

\bibitem{Harvey:2005ab}
J.~A. Harvey and S.~Jensen, ``Worldsheet instanton corrections to the
  {K}aluza-{K}lein monopole,''
  \href{http://dx.doi.org/10.1088/1126-6708/2005/10/028}{{\em JHEP} {\bfseries
  10} (2005) 028},
\href{http://arxiv.org/abs/hep-th/0507204}{{\ttfamily arXiv:hep-th/0507204}}.

\bibitem{Tong:2002rq}
D.~Tong, ``{NS}5-branes, {T}-duality and worldsheet instantons,'' {\em JHEP}
  {\bfseries 07} (2002) 013,
\href{http://arxiv.org/abs/hep-th/0204186}{{\ttfamily arXiv:hep-th/0204186}}.

\bibitem{Hull:2004in}
C.~M. Hull, ``A geometry for non-geometric string backgrounds,''
  \href{http://dx.doi.org/10.1088/1126-6708/2005/10/065}{{\em JHEP} {\bfseries
  10} (2005) 065},
\href{http://arxiv.org/abs/hep-th/0406102}{{\ttfamily arXiv:hep-th/0406102}}.

\bibitem{Hull:2006va}
C.~M. Hull, ``Doubled geometry and {T}-folds,''
  \href{http://dx.doi.org/10.1088/1126-6708/2007/07/080}{{\em JHEP} {\bfseries
  07} (2007) 080},
\href{http://arxiv.org/abs/hep-th/0605149}{{\ttfamily arXiv:hep-th/0605149}}.

\bibitem{Dabholkar:2005ve}
A.~Dabholkar and C.~Hull, ``Generalised {T}-duality and non-geometric
  backgrounds,'' \href{http://dx.doi.org/10.1088/1126-6708/2006/05/009}{{\em
  JHEP} {\bfseries 0605} (2006) 009},
\href{http://arxiv.org/abs/hep-th/0512005}{{\ttfamily arXiv:hep-th/0512005
  [hep-th]}}.

\bibitem{Sen:1997zb}
A.~Sen, ``{K}aluza-{K}lein dyons in string theory,''
  \href{http://dx.doi.org/10.1103/PhysRevLett.79.1619}{{\em Phys. Rev. Lett.}
  {\bfseries 79} (1997) 1619--1621},
\href{http://arxiv.org/abs/hep-th/9705212}{{\ttfamily arXiv:hep-th/9705212}}.

\bibitem{Howe:1985pm}
P.~S. Howe and G.~Sierra, ``Two-dimensional supersymmetric nonlinear sigma
  models with torsion,''
\href{http://dx.doi.org/10.1016/0370-2693(84)90736-6}{{\em Phys. Lett.}
  {\bfseries B148} (1984) 451--455}.

\bibitem{Duff:1989tf}
M.~J. Duff, ``Duality rotations in string theory,''
\href{http://dx.doi.org/10.1016/0550-3213(90)90520-N}{{\em Nucl. Phys.}
  {\bfseries B335} (1990) 610}.

\bibitem{Tseytlin:1990nb}
A.~A. Tseytlin, ``Duality symmetric formulation of string world sheet
  dynamics,''
\href{http://dx.doi.org/10.1016/0370-2693(90)91454-J}{{\em Phys. Lett.}
  {\bfseries B242} (1990) 163--174}.

\bibitem{Tseytlin:1990va}
A.~A. Tseytlin, ``Duality symmetric closed string theory and interacting chiral
  scalars,''
\href{http://dx.doi.org/10.1016/0550-3213(91)90266-Z}{{\em Nucl. Phys.}
  {\bfseries B350} (1991) 395--440}.

\bibitem{Maharana:1992my}
J.~Maharana and J.~H. Schwarz, ``Noncompact symmetries in string theory,''
  \href{http://dx.doi.org/10.1016/0550-3213(93)90387-5}{{\em Nucl. Phys.}
  {\bfseries B390} (1993) 3--32},
\href{http://arxiv.org/abs/hep-th/9207016}{{\ttfamily arXiv:hep-th/9207016}}.

\bibitem{Siegel:1993th}
W.~Siegel, ``Superspace duality in low-energy superstrings,''
  \href{http://dx.doi.org/10.1103/PhysRevD.48.2826}{{\em Phys. Rev.} {\bfseries
  D48} (1993) 2826--2837},
\href{http://arxiv.org/abs/hep-th/9305073}{{\ttfamily arXiv:hep-th/9305073}}.

\bibitem{Cremmer:1997ct}
E.~Cremmer, B.~Julia, H.~Lu, and C.~N. Pope, ``Dualisation of dualities. i,''
  \href{http://dx.doi.org/10.1016/S0550-3213(98)00136-9}{{\em Nucl. Phys.}
  {\bfseries B523} (1998) 73--144},
\href{http://arxiv.org/abs/hep-th/9710119}{{\ttfamily arXiv:hep-th/9710119}}.

\bibitem{Hull:2009sg}
C.~M. Hull and R.~A. Reid-Edwards, ``Non-geometric backgrounds, doubled
  geometry and generalised {T}-duality,''
  \href{http://dx.doi.org/10.1088/1126-6708/2009/09/014}{{\em JHEP} {\bfseries
  09} (2009) 014},
\href{http://arxiv.org/abs/0902.4032}{{\ttfamily arXiv:0902.4032 [hep-th]}}.

\bibitem{HackettJones:2006bp}
E.~Hackett-Jones and G.~Moutsopoulos, ``Quantum mechanics of the doubled
  torus,'' \href{http://dx.doi.org/10.1088/1126-6708/2006/10/062}{{\em JHEP}
  {\bfseries 10} (2006) 062},
\href{http://arxiv.org/abs/hep-th/0605114}{{\ttfamily arXiv:hep-th/0605114}}.

\bibitem{Berman:2007vi}
D.~S. Berman and N.~B. Copland, ``The string partition function in {H}ull's
  doubled formalism,''
  \href{http://dx.doi.org/10.1016/j.physletb.2007.03.007}{{\em Phys.Lett.}
  {\bfseries B649} (2007) 325--333},
  \href{http://arxiv.org/abs/hep-th/0701080}{{\ttfamily arXiv:hep-th/0701080
  [hep-th]}}.

\bibitem{Berman:2007xn}
D.~S. Berman, N.~B. Copland, and D.~C. Thompson, ``Background field equations
  for the duality symmetric string,''
  \href{http://dx.doi.org/10.1016/j.nuclphysb.2007.09.021}{{\em Nucl.Phys.}
  {\bfseries B791} (2008) 175--191},
  \href{http://arxiv.org/abs/0708.2267}{{\ttfamily arXiv:0708.2267 [hep-th]}}.

\bibitem{Berman:2007yf}
D.~S. Berman and D.~C. Thompson, ``Duality symmetric strings, dilatons and
  {O}(d,d) effective actions,''
  \href{http://dx.doi.org/10.1016/j.physletb.2008.03.012}{{\em Phys. Lett.}
  {\bfseries B662} (2008) 279--284},
\href{http://arxiv.org/abs/0712.1121}{{\ttfamily arXiv:0712.1121 [hep-th]}}.

\bibitem{Witten:1993yc}
E.~Witten, ``Phases of {$N=2$} theories in two dimensions,''
  \href{http://dx.doi.org/10.1016/0550-3213(93)90033-L}{{\em Nucl. Phys.}
  {\bfseries B403} (1993) 159--222},
\href{http://arxiv.org/abs/hep-th/9301042}{{\ttfamily arXiv:hep-th/9301042}}.

\bibitem{Mermin:1966fe}
N.~D. Mermin and H.~Wagner, ``Absence of ferromagnetism or antiferromagnetism
  in one- dimensional or two-dimensional isotropic {H}eisenberg models,''
\href{http://dx.doi.org/10.1103/PhysRevLett.17.1133}{{\em Phys. Rev. Lett.}
  {\bfseries 17} (1966) 1133--1136}.

\bibitem{Coleman:1973ci}
S.~R. Coleman, ``There are no {G}oldstone bosons in two-dimensions,''
\href{http://dx.doi.org/10.1007/BF01646487}{{\em Commun. Math. Phys.}
  {\bfseries 31} (1973) 259--264}.

\bibitem{Affleck:1980mp}
I.~Affleck, ``On constrained instantons,''
\href{http://dx.doi.org/10.1016/0550-3213(81)90307-2}{{\em Nucl. Phys.}
  {\bfseries B191} (1981) 429}.

\bibitem{Okuyama:2005gx}
K.~Okuyama, ``Linear sigma models of {H} and {KK} monopoles,''
  \href{http://dx.doi.org/10.1088/1126-6708/2005/08/089}{{\em JHEP} {\bfseries
  08} (2005) 089},
\href{http://arxiv.org/abs/hep-th/0508097}{{\ttfamily arXiv:hep-th/0508097}}.

\bibitem{Shapere:1988zv}
A.~D. Shapere and F.~Wilczek, ``Selfdual models with theta terms,''
\href{http://dx.doi.org/10.1016/0550-3213(89)90016-3}{{\em Nucl. Phys.}
  {\bfseries B320} (1989) 669}.

\bibitem{Giveon:1988tt}
A.~Giveon, E.~Rabinovici, and G.~Veneziano, ``Duality in string background
  space,''
\href{http://dx.doi.org/10.1016/0550-3213(89)90489-6}{{\em Nucl. Phys.}
  {\bfseries B322} (1989) 167}.

\bibitem{Witten:2009xu}
E.~Witten, ``Branes, instantons, and {T}aub-{NUT} spaces,''
  \href{http://dx.doi.org/10.1088/1126-6708/2009/06/067}{{\em JHEP} {\bfseries
  06} (2009) 067},
\href{http://arxiv.org/abs/0902.0948}{{\ttfamily arXiv:0902.0948 [hep-th]}}.

\bibitem{Hull:2009mi}
C.~Hull and B.~Zwiebach, ``Double field theory,''
  \href{http://dx.doi.org/10.1088/1126-6708/2009/09/099}{{\em JHEP} {\bfseries
  09} (2009) 099},
\href{http://arxiv.org/abs/0904.4664}{{\ttfamily arXiv:0904.4664 [hep-th]}}.

\bibitem{Hohm:2010jy}
O.~Hohm, C.~Hull, and B.~Zwiebach, ``Background independent action for double
  field theory,'' \href{http://dx.doi.org/10.1007/JHEP07(2010)016}{{\em JHEP}
  {\bfseries 07} (2010) 016},
\href{http://arxiv.org/abs/1003.5027}{{\ttfamily arXiv:1003.5027 [hep-th]}}.

\bibitem{Hohm:2010pp}
O.~Hohm, C.~Hull, and B.~Zwiebach, ``Generalized metric formulation of double
  field theory,'' \href{http://dx.doi.org/10.1007/JHEP08(2010)008}{{\em JHEP}
  {\bfseries 08} (2010) 008},
\href{http://arxiv.org/abs/1006.4823}{{\ttfamily arXiv:1006.4823 [hep-th]}}.

\end{thebibliography}\endgroup

\end{document}